\documentclass[a4paper,pre,reqno,superscriptaddress,twocolumn]{revtex4}
\usepackage{graphicx,color}
\usepackage{dcolumn}
\usepackage{epsfig}  
\usepackage[centertags]{amsmath}
\usepackage{amsfonts}
\usepackage{euscript}
\usepackage{amssymb}
\usepackage{amsthm}
\usepackage{newlfont}
\usepackage{lipsum}
\usepackage{mathtools}
\usepackage{mathrsfs}
\usepackage{subfigure}
\usepackage{todonotes,changes}

\newcommand{\opunit}{\text{1}\kern-0.22em\text{l}}

\newcommand{\ie}{\textit{i.e.}}

\newcommand{\id}{\textrm{d}}

\def\bea{\begin{eqnarray}}
\def\eea{\end{eqnarray}}
\def\ba{\begin{array}}
\def\ea{\end{array}}
\def\n{\nonumber}
\def\la{\langle}
\def\ra{\rangle}

\def\dr{D_R}

\begin{document}
 
\title{Active Brownian Motion in two-dimensions under Stochastic Resetting}
\author{Vijay Kumar}
\affiliation{Department of Chemical Engineering, Indian Institute of Science, Bengaluru, India}
\author{Onkar Sadekar}
\affiliation{Indian Institute of Science Education and Research, Homi Bhabha Road, Pashan, Pune, India}
\author{Urna Basu}
\affiliation{Raman Research Institute, C. V. Raman Avenue, Bengaluru, India}


\begin{abstract}
We study the position distribution of an active Brownian particle (ABP) in the presence of stochastic resetting in two spatial dimensions. We consider three different resetting protocols :  (I) where both position and orientation of the particle are reset,   (II) where only the position is reset, and (III) where only the orientation  is reset  with a certain rate $r.$ We show that in the first two cases the ABP reaches a stationary state. Using a renewal approach, we calculate exactly the  stationary marginal position distributions in the limiting cases when the resetting rate $r$ is much larger or much smaller than the rotational diffusion constant $\dr$ of the ABP. We find that, in some cases, for a large resetting rate, the position distribution diverges near the resetting point; the nature of the divergence depends on the specific protocol.
For the orientation resetting, there is no stationary state, but the motion changes from a ballistic one at short-times to a diffusive one at late times. We characterize the short-time non-Gaussian marginal position distributions using a perturbative approach. 
\end{abstract}

\maketitle

\section{Introduction}

Stochastic resetting refers to intermittent interruption and restart of a dynamical process. Introduction of such resetting mechanism to a stochastic process changes both static and dynamical properties of the system drastically \cite{EvansReview2019}. Study of resetting is relevant in a wide range of areas including search problems \cite{search1,search4,Arnab2017,Arnab2019}, population dynamics \cite{population1, population2}, computer science\cite{search2,search3},  and biological processes \cite{bio1, bio2, bio3}. The paradigmatic example of stochastic resetting is that of a Brownian diffusive particle which is reset to its initial position with some rate \cite{Brownian}. The presence of the resetting drives the system out of equilibrium, which leads to a lot of interesting behaviour including nonequilibrium steady states, dynamical transition in the  temporal relaxation and non-monotonic mean first passage time \cite{Brownian,Brownian2,MajumdarPRE2015}. Effect of resetting on various other diffusive processes have also been studied over the last decade \cite{Brownian3,absorption,highd,Mendez2016,Puigdellosas,Arnab2019_2,trap_reset, deepak,Experiment_reset, Arnab2015, potential, Prashant,Schehrreset2020}. A natural question that arises is what happens when resetting is introduced to a system where the underlying stochastic process  is `active' instead of passive diffusion.


Active processes refer to a class of dynamics which are intrinsically out of equilibrium due to self-propulsion~\cite{Romanczuk,soft,BechingerRev,Ramaswamy2017,Marchetti2017,motile2020}. Since the seminal work of Vicsek~\cite{Vicsek}, there has been a huge surge of interest in active matter systems which show a set of novel collective behaviour like flocking \cite{flocking1, flocking2}, clustering \cite{cluster1,cluster2,evans}, motility induced phase separation \cite{separation1, separation2, separation3,motile2020}. 
Theoretical attempts to understand the properties of active matter focuses on studies of simple yet analytically tractable models, like Run and Tumble particles (RTP), active Brownian particle (ABP) and their many variations \cite{BechingerRev}. 
In such models, the active nature of the dynamics emerges due to a coupling of the spatial motion with some internal `orientation' degree of freedom which itself evolves stochastically. The presence of an intrinsic time-scale associated with the internal orientation leads to a lot of interesting behaviour even at a single particle level which includes spatial  anisotropy and ballistic motion at short-times \cite{ABP2018, majumdarABP2020, Santra2020}, non-Boltzman stationary state and  clustering near the boundaries of the confining region \cite{Berke2008,Cates2009,Solon2015,Potosky2012,ABP2019,RTP_trap,Malakar2019} and unusual relaxation and first-passage properties \cite{RTP_free, ABP2018, Singh2019}. 


The first step to study the effect of resetting on active processes is to investigate the behaviour of a single active particle under stochastic resetting.
 Since active particles are characterized by both position and orientation degrees, the resetting can be defined in the phase space instead of position space, which opens up various possibilities regarding resetting protocols.  The presence of stochastic resetting introduces an additional time-scale given by the inverse of the resetting rate. For active particles, the interplay between the internal time-scale  and that of the resetting is expected to lead to a richer behaviour compared to its passive counterpart. 
Indeed, it has recently been shown that introduction of a stochastic resetting to the dynamics of an RTP leads to  non-trivial stationary distribution and first passage properties \cite{RTP_reset}. The first-passage properties of ABP and RTP under various resetting mechanisms have also been investigated recently \cite{Bressloff2020, Scacchi2017, Bressloff2020_2}.

\begin{figure*}
    \centering
    \includegraphics[width=17.5cm]{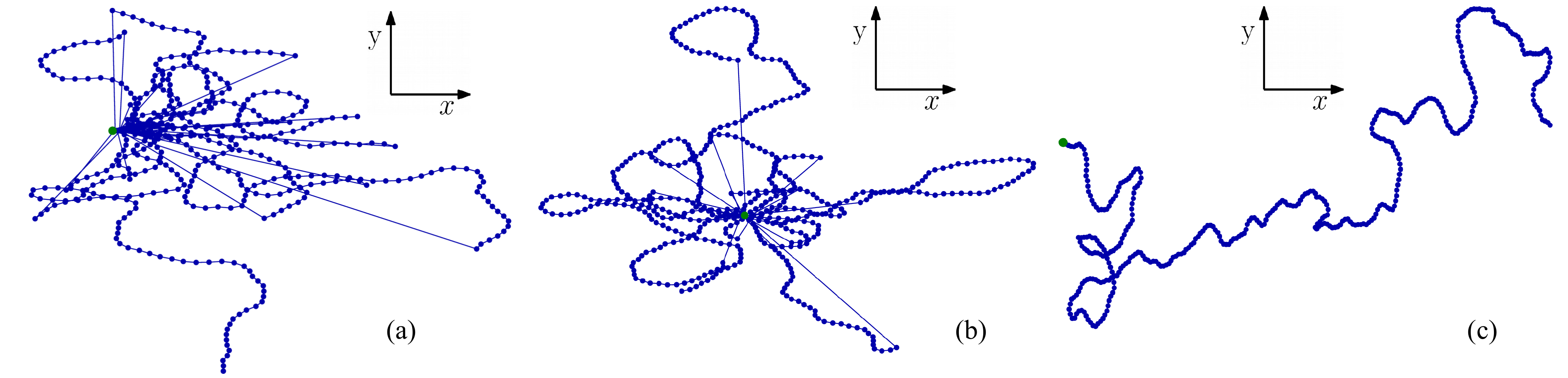}
    \caption{Typical trajectories of an ABP for the three different resetting protocols: position--orientation resetting (a), position resetting (b) and orientation resetting (c). In all the cases the particle starts from $x=0=y$ along $\theta=0.$ 
    }
    \label{fig:trajectory}
\end{figure*}

In this article we study the effect of stochastic resetting on active Brownian motion in two spatial dimension. An active Brownian particle (ABP) is an overdamped particle with an internal orientation which undergoes a rotational diffusion. Consequently, in two spatial dimension, an ABP is characterized by its position $(x,y)$ as well as its orientation $\theta.$  We study three different resetting protocols: (I) The position and the orientation of the particle are reset to their initial values with rate $r$, (II) only the position is reset, and (III) only the orientation is reset. 
In the first two cases, \ie, where the resetting protocol involves the resetting of the position, the particle position reaches a stationary state. We show that depending on whether the resetting rate $r$ is larger or smaller compared to the rotational diffusion constant $\dr,$ the stationary position distribution is very different.
We compute exactly the marginal position distributions  for the two liming scenarios, namely, $r \ll \dr$ and $r \gg \dr.$ It turns out that, for protocol I, the position distribution is strongly anisotropic for  $r \gg \dr;$ while for the protocol II, the distribution remains isotropic. Moreover, we show that, for large $r \gg \dr,$ in some cases, the stationary distribution diverges near the resetting position; the nature of the divergence depends on the  resetting protocol.

For purely orientational resetting, \ie, for protocol III, the particle does not reach a stationary state, but shows an anisotropic motion with a ballistic to diffusive crossover as time progresses.  We show that, at late times, the typical fluctuations of the position around its mean values are characterized by a Gaussian distribution. In the short-time regime, the position fluctuations are non-Gaussian; we adopt a perturbative method to compute the same for small values of the resetting rate.

In the next section we define the resetting protocols in details and present a brief summary of our results. Sections \ref{sec:protocolI} and \ref{sec:protocolII} are devoted to the study of the position-orientation resetting and position resetting cases, respectively.  The behaviour of the ABP under orientation resetting only is discussed in Sec.~\ref{sec:protocolIII}. We conclude with some general remarks in Sec.~\ref{sec:concl}.

\section{Model and Results} \label{sec:models}
        
Let us consider an active Brownian particle moving with a constant speed $v_0$ on a two-dimensional plane. Apart from the position coordinates $(x,y),$ the particle also has an internal degree of freedom, characterized by the orientation $\theta$, which itself undergoes a rotational Brownian motion. The Langevin equations describing this active Brownian motion are, 
\bea 
\dot x(t) &=& v_0 \cos \theta(t), \cr
\dot y(t) &=& v_0 \sin \theta(t), \cr
\dot \theta(t) &=& \sqrt{2 \dr}~ \eta (t), \label{eq:langevin}
\eea 
where $\eta(t)$ is a delta-correlated white noise and $\dr$ is the rotational diffusion constant. The coupling between the position and orientation degrees leads to the `active' nature of the motion. 
The activity, in turn, gives rises to various intriguing behaviour including 
non-trivial position distributions at short-times which crosses over to an effective diffusive behaviour at late-times. For the sake of completeness, a brief review of the behaviour of ordinary ABP is provided in Appendix \ref{app:free_abp}.


\begin{figure*}
\centering
   \includegraphics[width=17.5cm]{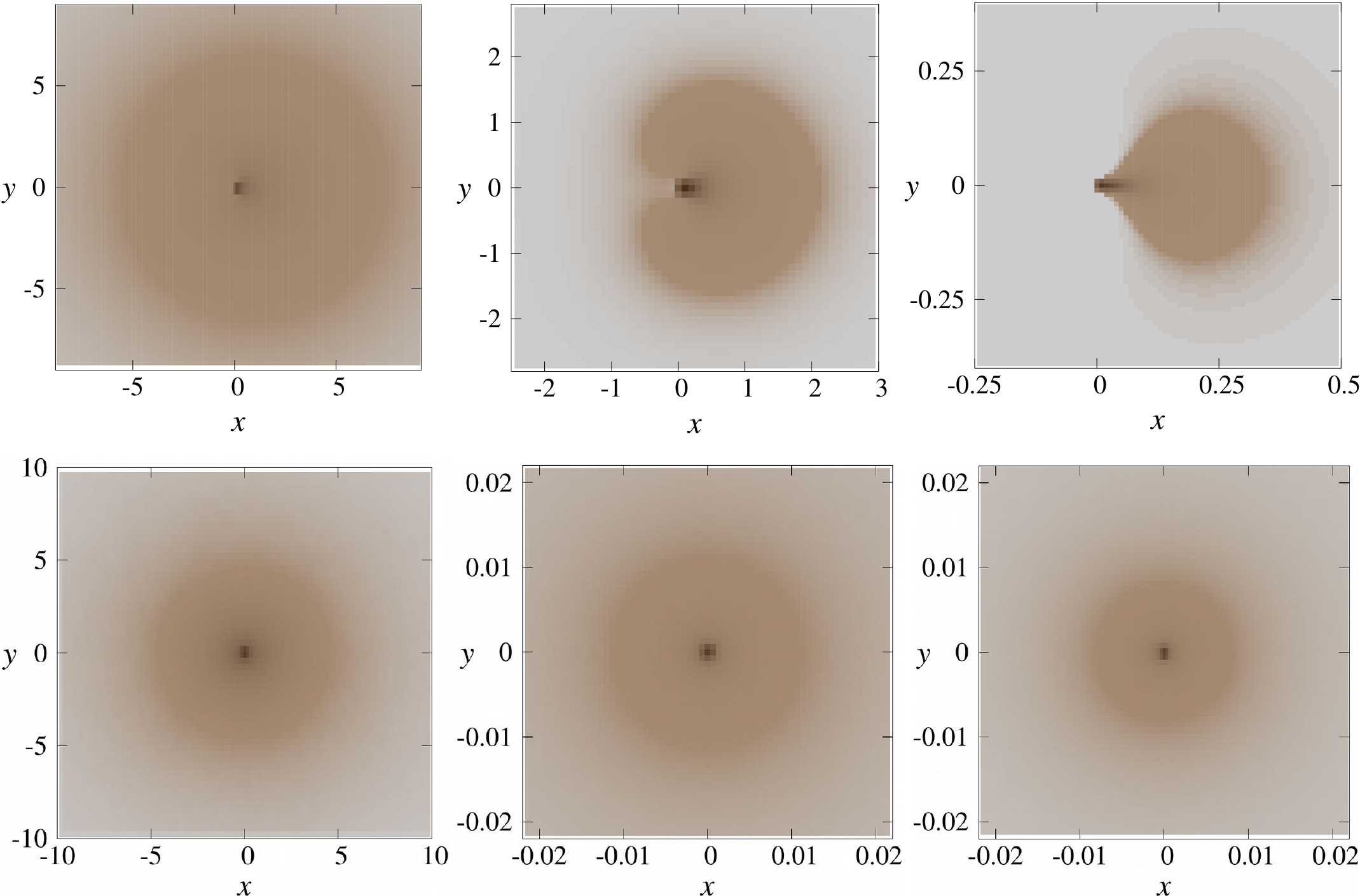}
\caption{ Plot of the stationary position probability distribution $P_\text{st}(x,y)$ in the $x-y$ plane for resetting protocols I (upper panel) and II (lower panel). The darker region corresponds to higher value of the probability density.  The left, middle and right columns correspond to $r=0.01,\dr=1,$ $r=1,\dr=1$ and $r=10,\dr=1$ respectively, with $v_0=1$ for all the cases. }
    \label{fig:2d_dist}
\end{figure*}

In this article we study the effect of stochastic resetting on the dynamics of such an active Brownian particle. Since the ABP is characterized by both the position and orientation degrees, the resetting might affect both these degrees. In the following, we focus on three different resetting protocols. 

\begin{itemize}
 \item [I.]{\bf Resetting of the position and orientation}: In this case, the position of the particle, along with its orientation is reset to the corresponding initial values with rate $r.$ We assume that the particle starts from the origin, oriented along the $x$-axis, so that, at any time $t,$ the particle is reset to  $x=0=y=\theta$, with rate $r.$  The system reaches a stationary state in the long-time limit. We investigate the stationary marginal position distributions as well as the time evolution of the moments of the position.

\item[II.]{\bf Resetting of the position}: In the second scenario we reset the position of the particle to the origin with rate $r,$ but the orientation is not affected -- it evolves as a free Brownian motion. In this case also the ABP reaches a stationary state. We characterize the moments and the stationary marginal position distributions. 

\item[III.] {\bf Resetting of the orientation}: In this  scenario, only the orientation $\theta$ is reset with rate $r,$ the position degrees are not affected. In this case the position distribution does not reach a stationary state; we study the short-time and long-time limiting behaviour of the marginal distributions along with the position moments. 
\end{itemize} 

Figure~\ref{fig:trajectory} shows typical trajectories of an ABP in the presence of these three resetting protocols.  In the first case (protocol I), the particle preferably visits the right half-plane $x>0$ because of the resetting of the orientation while for protocol II, the motion looks more isotropic. For protocol III, the particle runs along the $x$-axis, away from the origin. 

In the absence of resetting, the active Brownian particle shows an interesting dynamical crossover depending on the value of the rotational diffusion constant $\dr.$ Starting from the origin, and with $\theta=0,$ at short-times $t \ll \dr^{-1},$ the motion is strongly non-diffusive and the position distribution remains anisotropic with the variance along the $x$ and $y$ directions showing very different temporal growths \cite{ABP2018,majumdarABP2020}. 
At long times $t \gg \dr^{-1},$ however, the motion becomes diffusive and the typical position fluctuations become Gaussian in nature, with only the tails retaining signatures of activity \cite{ABP2019}. The presence of stochastic resetting introduces another timescale $r^{-1}$, \ie, the inverse of the resetting rate.  We expect that the interplay of the two time scales $\dr^{-1}$ and $r^{-1}$ would lead  to a rich behaviour for ABP under resetting. 


Figure~\ref{fig:2d_dist} illustrates  the qualitative nature of the 2D stationary position distribution $P_\text{st}(x,y)$ for resetting protocols I (upper panel) and II (lower panel). The left column shows the distribution for $r \ll \dr,$ where, in both cases, the distribution is isotropic. The middle column shows the same for $r \sim \dr,$ where for protocol I the distribution becomes anisotropic.
For protocol II, the distribution remains isotropic, but the width decreases as $r$ is increased.  The anisotropy becomes stronger for protocol I as $r$ is increased, as can be seen from the right panel ($r \gg \dr$). The anisotropy for the position-orientation resetting arises due to the fact that after each resetting, the orientation is brought back to $\theta=0,$ and the particle restarts motion along the $x$-axis. On the other hand, for protocol II, \ie, when the resetting does not affect the orientation of the particle, the stationary distribution remains isotropic for all values of $r$ and $\dr.$

\begin{figure*}
\centering
   \includegraphics[width=17.5cm]{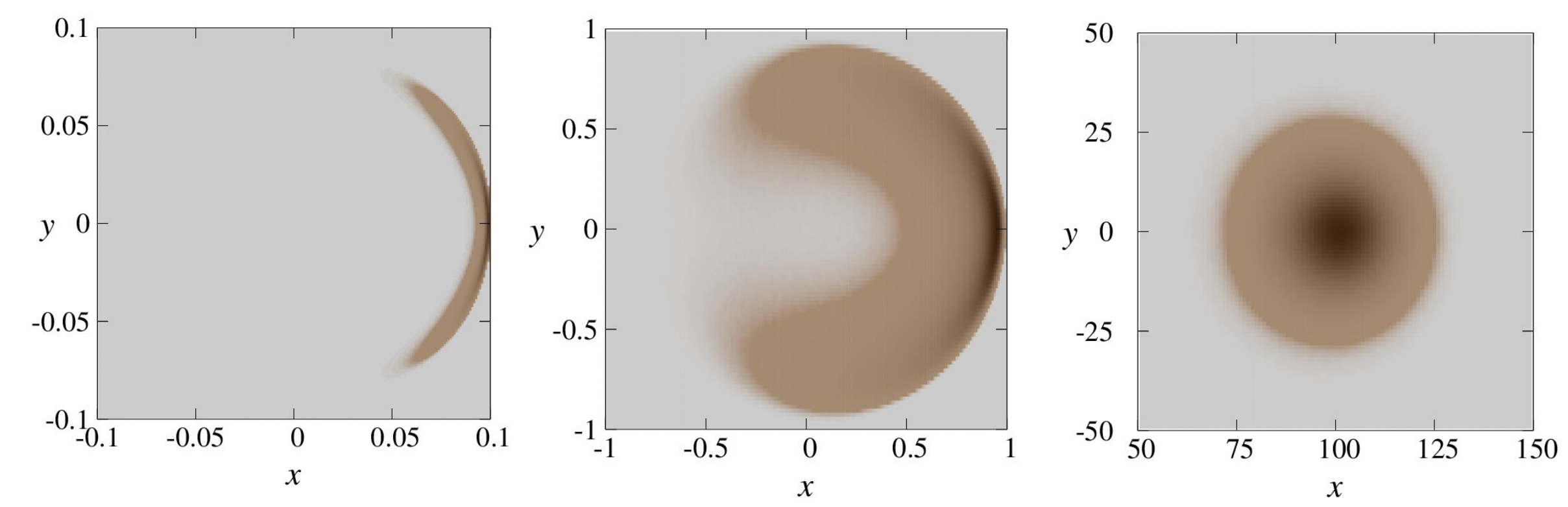}
    \caption{Plot of position distribution $P(x,y,t)$ in the $x-y$ plane  for orientation resetting for different values of time $t=0.1$ (left), $t=1$ (middle) and $t=200$ (right). Darker regions correspond to higher values of the probability.  Here $r=1=\dr$ and $v_0=1.$}
    \label{fig:2d_dist_III}
\end{figure*}

As mentioned already, for protocol III, \ie, for orientational resetting, the particle position does not reach a stationary state. In this case, the nature of the motion changes from ballistic at short-times to diffusive at late times. This is qualitatively illustrated in Fig.~\ref{fig:2d_dist_III} where $P(x,y,t)$ is shown for three different values of time. At short-times (left panel) the distribution remains strongly anisotropic, similar to the free ABP case. The anisotropy decreases as time is increased (middle panel), ultimately reaching a Gaussian-like distribution at late times $ t \gg (r+\dr)^{-1},$ as we will demonstrate later.  

Before going to the details of the computations we first present a brief summary of our results. 

\begin{itemize}
\item We show that the position distribution reaches a stationary state if the resetting protocol involves changing the position directly, \ie, for protocols I and II. We study the corresponding stationary marginal distributions and show that depending on whether the time scale $r^{-1}$ associated with resetting is larger or smaller than the inherent rotational time-scale $\dr^{-1}$ of the ABP, the stationary distribution has very different forms.

\item For protocol I, the stationary distribution is strongly anisotropic for $r \gg \dr.$ In this case, the $x$-marginal distribution falls off exponentially for large $x>0,$ while approaching a finite value near the origin [see Eq.~\eqref{eq:tailx_proI_large_r}]. The $x<0$ region remains unpopulated. The $y$-distribution, however, turns out to be symmetric and shows an algebraic divergence $|y|^{-1/3}$ near the origin, while decaying as a compressed exponential for large $|y|$ [see Eqs.~\eqref{eq:Pst_pos_th_ysmall} and \eqref{eq:Pst_pos_th_largey}]. 

For $r \ll \dr,$ on the other hand, the anisotropy disappears, both $x$ and $y$-marginal distributions attain exponential forms; see Eqs.~\eqref{eq:Px_proI} and \eqref{eq:Py_rsmall}.  

\item For protocol II, the stationary distribution remains isotropic for all parameter values. In this case, for $r \ll \dr $ the distribution is exponential in nature, similar to protocol I; see Eq.~\eqref{eq:Pst_pos_large_D}. 

For $r \gg \dr,$ the distribution becomes independent of $\dr,$ and shows a log-divergence near the origin [see Eq.~\eqref{eq:pos_log_div}] while decaying as an exponential for large $|x|$ [see Eq.~\eqref{eq:pos_large_x}]. 

\item For the protocol III, the position distribution does not reach a stationary state. We show that at late times, the particle shows a diffusive behaviour; the typical position fluctuations are characterized by  Gaussian distributions in this limit. We compute the corresponding effective diffusion constants, which turn out to be different for $x$ and $y$ components, signaling presence of an anisotropy even at late times.

At short times, for this protocol, the motion remains ballistic. We compute the position distribution for small values of the resetting rate $r$, using a perturbative approach. The perturbative corrections corresponding to $x$ and $y$-distributions are obtained in Eqs.~\eqref{eq:P1x} and \eqref{eq:P1y}, respectively. 

\end{itemize}

In the following sections we study the three protocols separately and characterize  the fluctuations of the position by computing the moments and marginal distributions.

\section{ABP with position and orientation resetting}\label{sec:protocolI}

The simplest resetting protocol is when both the position and the orientation of the particle are reset to their initial values, with rate $r.$ This is referred to as protocol I in Sec.~\ref{sec:models}. For the sake of simplicity we assume that the ABP starts at the origin $x=y=0,$ oriented along the $x$-axis, \ie, with $\theta=0$ at time $t=0.$ Then, at any time $t,$ the ABP is reset to $x=y=0=\theta$ with rate $r;$ between two consecutive resetting events, the particle position evolves according to the Langevin equations \eqref{eq:langevin}. In the following we refer to this resetting protocol as `position-orientation reset'.

\begin{figure}
    \centering
    \includegraphics[width=8.8 cm]{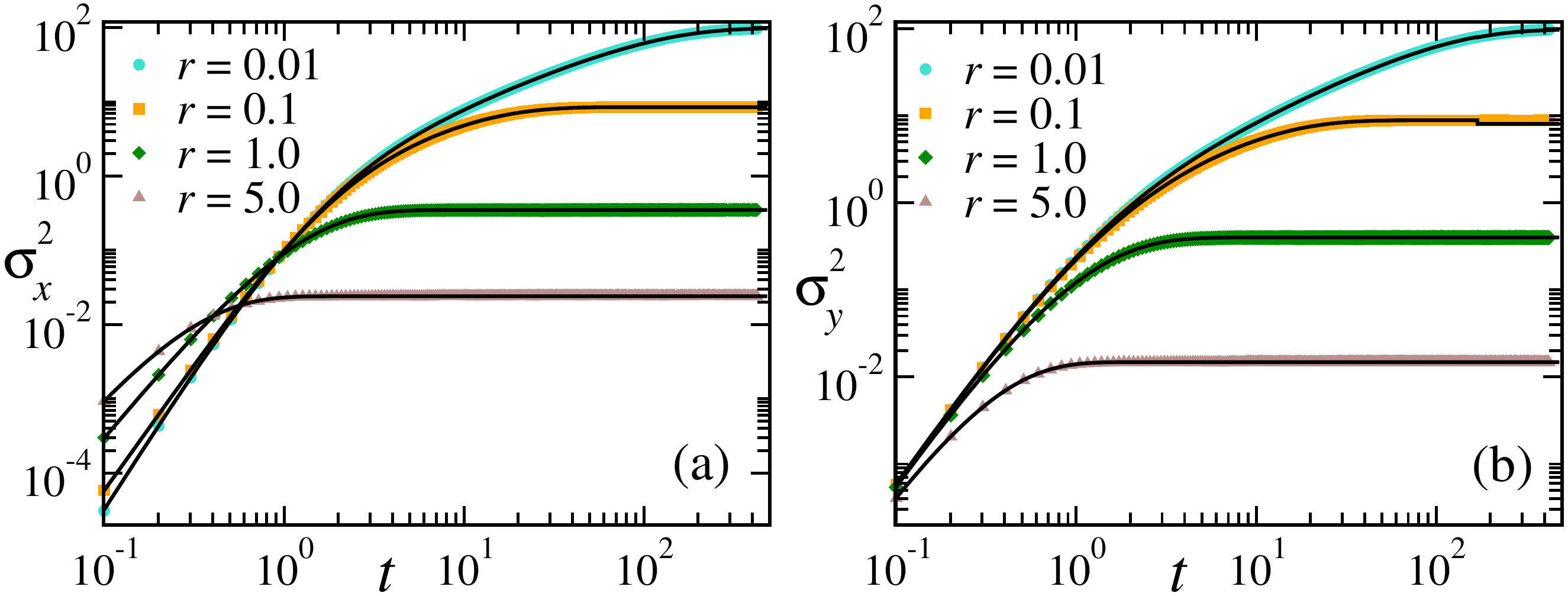}
    \caption{Position-orientation resetting: Mean squared displacements of $x$ and $y$ components as a function of time for $\dr=1$ and  different values of the resetting rate $r.$ Symbols represent the data from simulations while the solid black curves indicate the analytical predictions from Eqs.~\eqref{eq:varxt_pos_th} and \eqref{eq:varyt_pos_th}. Here $v_0=1.$}
    \label{fig:Moments_p_reset}
\end{figure}

We are interested in the position distribution $P(x,y,t) = \int \id \theta ~\cal  P(x,y,\theta,t)$  where $\cal  P(x,y,\theta,t)$ denotes the probability that the particle is at the position $(x,y)$ with orientation $\theta,$ at time $t.$
It is straightforward to write a renewal equation for $\cal  P(x,y,\theta,t)$ which reads, 
\bea
\cal  P(x,y,\theta,t) = e^{-rt} \cal P_0(x,y,\theta,t) + r \int_0^t \id s~ e^{-r s} \cal P_0(x,y,\theta, s), \n
\eea
where $\cal P_0(x,y,\theta,t)$ denotes the probability that in the absence of resetting, the ABP is at a position $(x,y)$ with orientation $\theta$ at time $t,$ starting from $x=0=y=\theta.$ Here the first term corresponds to the situation when there are no resetting events up to time $t$ and the second term corresponds to the probability that the last resetting event occurred at a time $t-s.$

A corresponding renewal equation for the position distribution is obtained by integrating over the orientation $\theta,$ 
\bea   
P(x,y,t) &=& e^{-r t} P_0(x,y,t) + r \int_0^t \id s~ e^{-r s} P_0(x,y,s).\quad \label{eq:renewal_pos_th}
\eea 
From this renewal equation, the position distribution can, in principle, be calculated for any time $t$, if the free ABP distribution is known. Unfortunately,  no closed form for the full distribution $P_0(x,y,t)$ of an ABP is known so far.
However, the short-time and long-time marginal position distributions are known explicitly \cite{ABP2018, ABP2019}, and  in this section we use these to investigate the effect of the  position-orientation resetting on an ABP using the renewal equation \eqref{eq:renewal_pos_th}.

\subsection{Moments}\label{sec:moments_protocolI}

To get an idea about how the presence of the position-orientation resetting affects the dynamical behaviour of the ABP, let us first look at the moments of the position coordinates. It is straightforward to see that, in the presence of the resetting, the moments would also satisfy a renewal equation similar to Eq.~\eqref{eq:renewal_pos_th}. For example, by multiplying both sides by $x^n$ and integrating over $x$ and $y,$ we get, 
\bea
\la x^n(t) \ra = e^{-r t} \la x^n(t) \ra_0 + r \int_0^t \id s ~e^{-r s} \la x^n(s) \ra_0, \label{eq:xn_pos_th}
\eea
where $\la x^n(t) \ra_0$ denotes the $n^{th}$ moment of the $x$-component of the position in the absence of the resetting which can be calculated explicitly for any $n$ \cite{ABP2018, Shee2020}. The renewal equation for $\la y^n(t) \ra$ also has a similar form. In the following we calculate explicitly the first two moments of $x$ and $y$-components using the known expressions for the same for free ABP (see Appendix~\ref{app:free_abp}). 

\begin{figure*}
\includegraphics[width=11.85 cm]{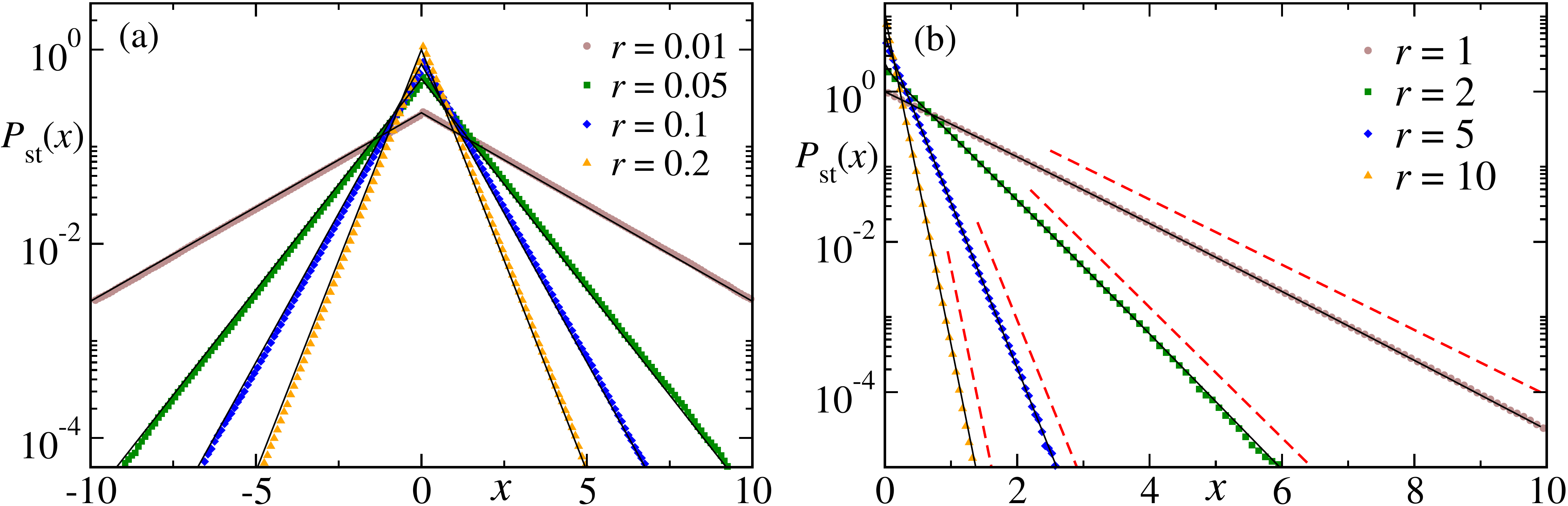} \hspace*{0.1cm}\includegraphics[width=5.65 cm]{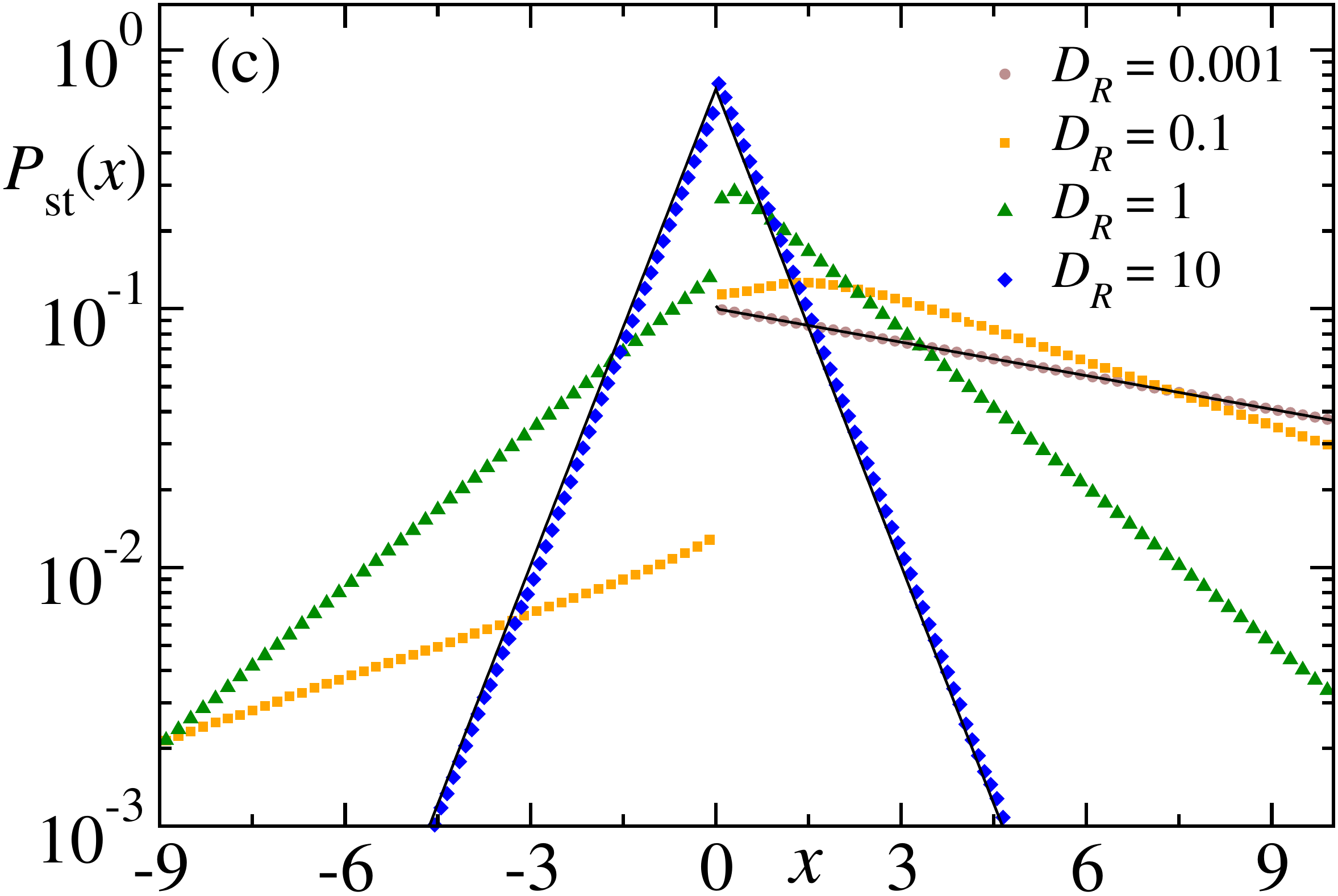} 
    \caption{Stationary $x$-marginal distribution for position-orientation resetting: (a) Plot of $P_\text{st}(x)$ versus $x$ for different values of $r$ in the regime $r \ll \dr$ and a fixed $\dr=10.$  (b) Similar plot in the regime $r \gg \dr$ with  $\dr=0.01.$  (c) The crossover between the two regimes for a fixed value of $r=0.1$ and different values of $\dr.$  The solid black lines indicate the analytical predictions [see Eq.~\eqref{eq:Px_proI} for (a) and Eq.~\eqref{eq:Px_proI_large_r} for (b)], the symbols show the data from numerical simulations and the red dashed line in (b) shows the exponential trend at large values of $x$. For all the plots  $v_0=1.$  }
    \label{fig:pdf_pos_th_reset_x}
\end{figure*}

Let us first look at the time-evolution of  the average position. Using Eq.~\eqref {eq:xn_pos_th} for $n=1$ along with Eq.~\eqref{eq:xt_yt_0}, we get,
\bea 
\la x(t) \ra & =& \frac{v_0}{r+\dr} (1- e^{-(r+\dr)t}),\label{eq:xt_pos_th} 
\eea   
while $\la y(t)\ra=0$ at all times. Here we see the first evidence of a new time-scale emerging as a result of the presence of the resetting. Clearly, at short-times, \ie, for $t \ll (r+\dr)^{-1},$ the particle moves along $x$-axis with an effective velocity $v_0$ which is reminiscent of the free ABP. On the other hand, at late-times the particle reaches a stationary position which comes closer to the origin as the resetting rate $r$ is increased. 
Next, we calculate the second moments using Eq.~\eqref {eq:xn_pos_th} with $n=2$ and Eq.~\eqref{eq:ABP_x2t_y2t}. The resulting exact (and long) expressions are provided in  Appendix~\ref{app:pos-th_rABP}. Here we explore the behaviour of the mean squared displacement (MSD) $\sigma_x^2(t)= \la x^2(t) \ra- \la x(t)\ra^2$ and $\sigma_y^2(t)= \la y^2(t) \ra$ in the short-time and the long-time regimes. At short-times,  we have,
\bea 
\sigma_x^2(t)&=&\frac{ v_0^2}{3} r t^{3}+\frac{v_0^2}{12}(4\dr^2-5\dr r-4r^2)t^{4}+O(t^5),\cr 
\sigma_y^2(t) &=&\frac{2 v_0^2}{3} \dr t^{3}-\frac{v_0^2}{6} \dr(5\dr+3r) t^{4}+O(t^5). 
\eea 
It is interesting to compare this short-time behaviour with that of ABP in the absence of resetting. Starting from the origin, oriented along the $x$-axis, for the ordinary ABP, the MSD along the $x$-direction grows $\sim t^4$ while along $y$, it shows a $t^3$ temporal growth. In the presence of the position-orientation resetting, however, we see that  both $\sigma_{x,y}^2$ grow as $t^3;$ while the resetting changes the leading order behaviour of the MSD along $x$-direction, it does not affect the same for the MSD along $y.$
      
At long-times, the particle is expected to reach a stationary state, and the MSD  does not depend on the time anymore,
\bea  
 \lim_{t \to \infty} \sigma_x^2 &=& \frac{v_0^2(4\dr^2+2 r \dr+r^2)}{r(r+\dr)^2(r+4\dr)},\cr
\lim_{t \to \infty} \sigma_y^2 &=&  \frac{4 v_0^2 \dr}{r(r+\dr)(r+4\dr)}. \label{eq:vary_pos_th}
\eea
It is to be noted that, the stationary values of the MSD are different for $x$ and $y$-components, indicating that the anisotropy survives. This is not surprising as the resetting to $\theta=0$ introduces strong anisotropy at each epochs.  Figure~\ref{fig:Moments_p_reset} show plots of $\sigma_x^2$ and $\sigma_y^2$ as functions of time $t$ for different values of $r;$ as expected, the MSD saturates faster to its stationary value with increasing $r$.

\subsection{Marginal $x$-distribution}

Let us consider the marginal $x$-distribution in the presence of position-orientation resetting. It satisfies a renewal equation obtained by integrating Eq.~\eqref{eq:renewal_pos_th} over $y,$
\bea
P(x,t) = e^{-r t} P_0(x,t) + r \int_0^t \id s~ e^{-r s} P_0(x,s),
\eea
where $P_0(x,s)$ denotes the $x$-marginal distribution in the absence of the resetting.  Note that for the sake of simplicity we use the same letter $P$ for both the $1-d$ and $2-d$ position distributions.

At late-times $t \to \infty,$ the particle position is expected to reach a stationary state. We concentrate on the stationary position distribution, which is given by,
\bea  
P_\text{st}(x) = r \int_0^\infty \id s~ e^{-rs} P_0(x,s).
\label{eq:p_reset_xst}
\eea 
As mentioned already, no closed form expressions are available for $P_0(x,t).$ 
However, the short-time ($ t \ll \dr^{-1}$) and long-time ($ t \gg \dr^{-1}$) behaviour of $P_0(x,t)$ are known separately \cite{ABP2018,ABP2019}.  In the following we show that, these short-time and long-time behaviour can be used to calculate the distribution in the  presence of  resetting in some cases. To this end, let us first recast Eq.~\eqref{eq:p_reset_xst}  as,
\bea 
P_\text{st}(x) = r \int_0^\infty \id u~ e^{-u} P_0(x,u/r).
\label{eq:p_reset_xst_scaled}
\eea 
Because of the presence of the $e^{-u}$ factor, the dominating contribution to the integral comes from $u \sim O(1).$ Then, depending on whether $u/r$ is large or small compared to $\dr^{-1}$, the dominant contribution comes from the large or short-time regime of the free ABP distribution. In the following, we discuss the two limiting cases separately. \\

\noindent {\bf Small resetting rate ($r \ll \dr$):} In this case the typical interval between two consecutive resetting events $r^{-1}$ is longer than the rotational time-scale $\dr^{-1},$ and the particle  evolves as a free ABP for a long-time   before being reset to the origin. Consequently, the  dominant contribution to the integral in Eq.~\eqref{eq:p_reset_xst_scaled} comes from the regime, $\frac u r \gg \dr^{-1}.$ In other words, we can use the long-time distribution of free ABP in Eq.~\eqref{eq:p_reset_xst} to compute the distribution in the presence of resetting. It has been shown that for $t \gg \dr^{-1},$ the free ABP distribution admits a large-deviation form,
\bea
P_0(x,t) \sim \exp{\left[- \dr s \Phi \left(\frac x{v_0 t} \right) \right]}, \label{eq:large_dev_x}
\eea
where the large deviation function $\Phi(z) = \frac {z^2}2 + O(z^4)$ \cite{ABP2019}. We are particularly interested in the typical fluctuations around $x=0,$ and it suffices to take the leading term, which, when normalized, leads to a Gaussian distribution,
\bea
P_0(x,t) = \sqrt{\frac{\dr}{2 \pi v_0^2 t}}\, \exp{\bigg[- \frac{\dr x^2}{2 v_0^2 t}\bigg]}.\label{eq:P0xs}
\eea
Substituting the above equation in Eq.~\eqref{eq:p_reset_xst} and performing the integral over $s,$ we get an exponential stationary distribution in the presence of resetting, 
\bea 
P_\text{st}(x) = \frac 1{v_0} \sqrt{\frac {r \dr}2} \exp{\left[-\sqrt{2r \dr} \frac{|x|}{v_0} \right]}. \label{eq:Px_proI}
\eea 
This distribution is symmetric around $x=0,$ and for large $x,$ falls faster as either $r$ or $\dr$ is increased. Figure ~\ref{fig:pdf_pos_th_reset_x} shows a plot of the predicted $P_\text{st}(x)$ for different (small) values of $r$ along with the same obtained from numerical simulations; an excellent match confirms that the prediction \eqref{eq:Px_proI} is valid for a substantial range of $r.$ \\


\noindent {\bf Large resetting rate  ($r \gg \dr$):}  In this case, the typical interval between two resetting events is much smaller compared to the rotational diffusion time-scale of the free ABP dynamics. Consequently, most trajectories evolve for a short-time before being reset to the origin. In other words, the  dominant contribution to the integral \eqref{eq:p_reset_xst} comes from the short-time regime of free ABP. It has been shown that, at short-times $t \ll \dr^{-1},$ the $x$-marginal distribution is given by a scaling form,
 \bea
 P_0(x,t)= \frac 1{v_0 \dr t^2} f_x\left(\frac{v_0 t -x}{v_0 \dr t^2}\right), \quad \text{for}~ x \le v_0t.
 \eea
Here the scaling function $f_x(u)$ is given by the sum of an infinite series. The explicit form of $f_x(u)$ is known and quoted in Appendix \ref{app:free_abp}; we use that to calculate $P_\text{st}(x)$ using Eq.~\eqref{eq:p_reset_xst}. Note that as $P_0(x,s)$ defined only in the regime $x \le v_0s,$ the lower limit of the integral becomes $s= x/v_0.$ This integral can be computed explicitly and yields a sum of exponentials,
\bea
P_\text{st}(x) &=& \frac{r \sqrt{\dr}}{2 v_0} \sum_{k=0}^{\infty} (-1)^k \frac{(4k+1)}{2^{2k}} \left({2 k \atop k}\right) \frac{\sqrt{a_k}+\sqrt{a_k+r}}{\sqrt{a_k(a_k+r)}}\cr
&& \exp{\Bigg[-\bigg(\sqrt{a_k}+\sqrt{a_k+r}\bigg)^2\frac x{v_0} \Bigg]}, \label{eq:Px_proI_large_r}
\eea
with $a_k= (4k+1)^2\dr/8.$  Note that, this expression is valid for $x>0.$ In fact, $x<0$ is not populated in this case, giving rise to a strong anisotropy, in contrast to the small $r$ case. Figure \ref{fig:pdf_pos_th_reset_x}(b) compares the analytical prediction \eqref{eq:Px_proI_large_r} with $P_\text{st}(x)$ obtained from numerical simulations for large values of $r$ which show perfect agreement. 

To understand the asymptotic behaviour for large $x,$ we note that for large $x,$ the exponential term with the smallest coefficient, \ie, with $k=0,$  would contribute. Hence, we expect that the tail of the distribution will have the form,
\bea
P_\text{st}(x) &\simeq& \frac{r}{v_0} \frac{\sqrt{2}\left(\sqrt{\dr}+\sqrt{\dr+8}\right)}{\sqrt{\dr+8}} \cr
&\times & \exp{\left[- \frac x{4 v_0}\bigg(4r+\dr + \sqrt{\dr(\dr+8r)}\bigg)\right]}.\qquad \label{eq:tailx_proI_large_r}
\eea
The exponential tails predicted in Eq.~\eqref{eq:tailx_proI_large_r} are indicated by red dashed lines in Fig.~\ref{fig:pdf_pos_th_reset_x}(b).

To explore how the  stationary distribution looks for intermediate values of $r,$ we take recourse to numerical simulations. Figure \ref{fig:pdf_pos_th_reset_x}(c) shows a plot of the same for different values of $r$ which shows the crossover from the asymmetric (one sided exponential for $x>0$) to the symmetric (exponential decay on both sides) distribution. We see that as $\dr$ is increased, the $x<0$ region starts to become populated, although the distribution remains strongly asymmetric, as indicated by the discontinuity across $x=0.$ The asymmetry disappears only for very large $\dr \gg r.$ 



\begin{figure}
\includegraphics[width=8 cm]{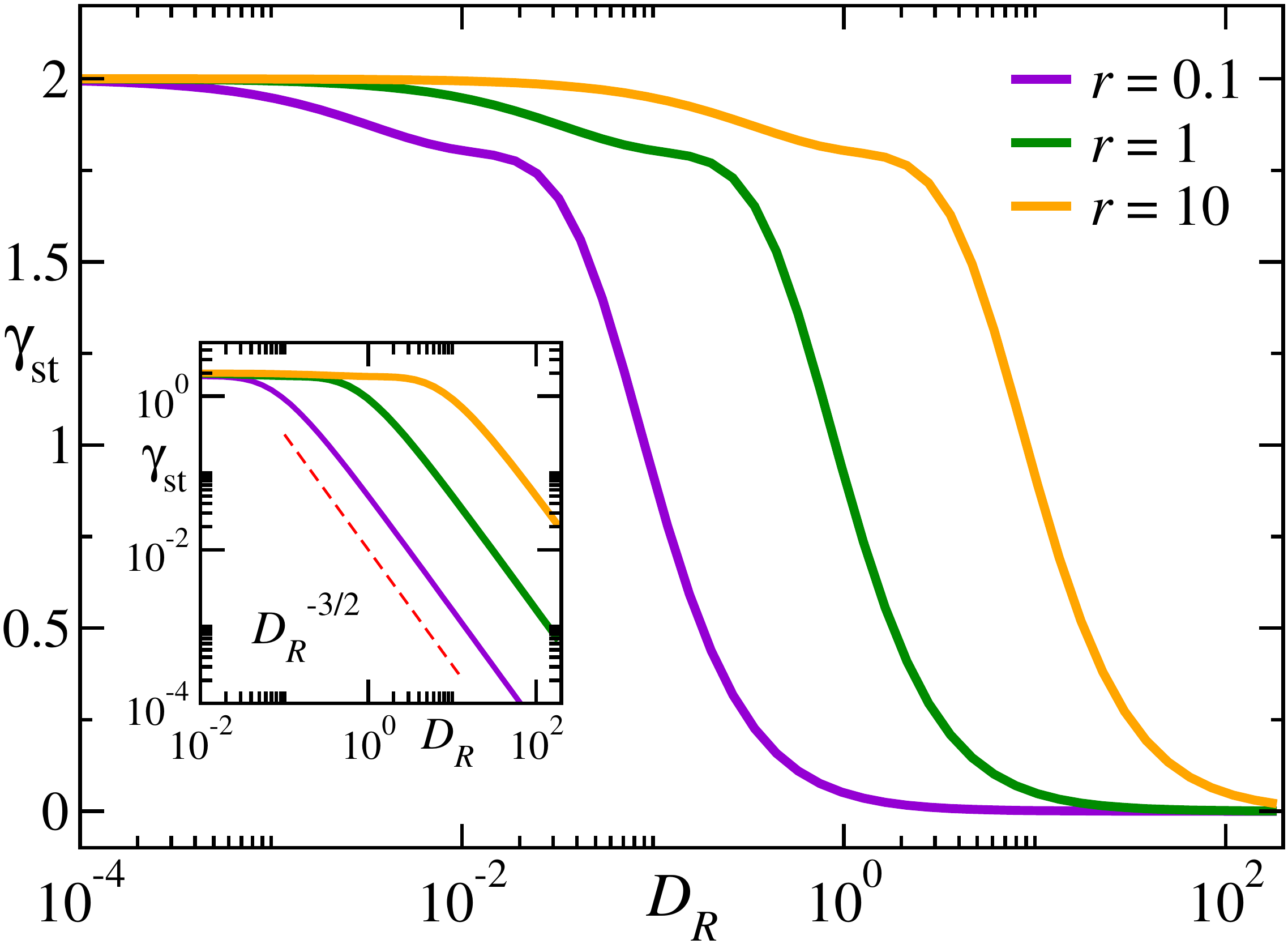}
\caption{Position orientation resetting: Plot of the steady state skewness $\gamma_\text{st}$ as a function of $\dr$ for different values of $r$ [see Eq.~\eqref{eq:skew_lt}]. For small values of $\dr \ll r$  $\gamma_{\text{st}} \rightarrow 2.$ The algebraic decay for large $\dr \gg r$ is shown in the inset [see Eq.~\eqref{eq:skew_limit}]. \label{fig:skew_pos_th}}
\end{figure}

To understand this crossover from a strongly asymmetric to symmetric behaviour of $P_\text{st}(x)$, we compute the skewness of $P(x,t)$, 
\bea
\gamma(t) =\frac{\la x^3(t) \ra -3 \la x(t) \ra  \sigma_x^2(t) -\la x(t) \ra^3}{\sigma_x^3(t)}.
\eea
The third moment of $x(t)$ can be calculated using Eq.~\eqref{eq:xn_pos_th} and Eq.~\eqref{eq:x3t_ABP}. The explicit expression for $\la x^3(t) \ra$ is provided in  Appendix \ref{app:pos-th_rABP}. However, since we are interested in the stationary distribution, it suffices to look at the stationary limit of the skewness $\gamma_{\text{st}}=\lim_{t \rightarrow \infty} \gamma(t).$
Substituting the expressions for moments and then taking the long-time limit we get,
\bea
\gamma_{\text{st}}=\frac{2 r \sqrt{r (4 \dr + r)} ~ (30 \dr^2 +7 \dr r +r^2)}{(9 \dr + r )~(4 \dr^2+ 2 \dr r+ r^2)^{3/2}}.\label{eq:skew_lt}
\eea
Figure~\ref{fig:skew_pos_th} shows a plot of $\gamma_{\text{st}}$ as a function of $\dr$ for a set of values of $r$. From Eq.~\eqref{eq:skew_lt} it is clear that for small values of $\dr \ll r,$   $\gamma_\text{st} \to 2,$ which indicates a strongly asymmetric distribution, as seen in Fig.~\ref{fig:pdf_pos_th_reset_x}(b). On the other hand, for  large values of $\dr,$ we have, 
\bea
 \gamma_{\text{st}} \simeq \frac{5}{3}\bigg(\frac{r}{\dr}\bigg)^{3/2}. \label{eq:skew_limit}
\eea
Hence, in the limit $r \ll \dr,$  the symmetric distribution ($\gamma_\text{st}=0$) is approached with an algebraic decay. 


\subsection{Marginal $y$-distribution}

The anisotropic nature of the position distribution, as seen in Fig.~\ref{fig:2d_dist}, indicates that the marginal distribution along $y$-direction is very different than the same along $x$-direction, at least for $r \ge \dr.$ In this section we investigate the behaviour of the marginal $y$-distribution in the presence of position-orientation resetting.  

The marginal distribution $P(y,t)$ satisfies a  renewal equation similar to the $x$-component,
\bea
P(y,t) = e^{-r t} P_0(y,t) + r \int_0^t \id s~ e^{-r s} P_0(y,s),
\eea
where $P_0(y,t)$ denotes the marginal distribution of ABP in the absence of resetting. 
Once again, we focus on the stationary distribution, and use the known short-time and long-time behaviours of the $P_0(y,t)$ to compute the position distribution in the presence of resetting. As before, we consider the two limiting cases where the resetting rate is much larger and smaller than the rotational diffusion constant. \\

\begin{figure*}
\includegraphics[width=11.85 cm]{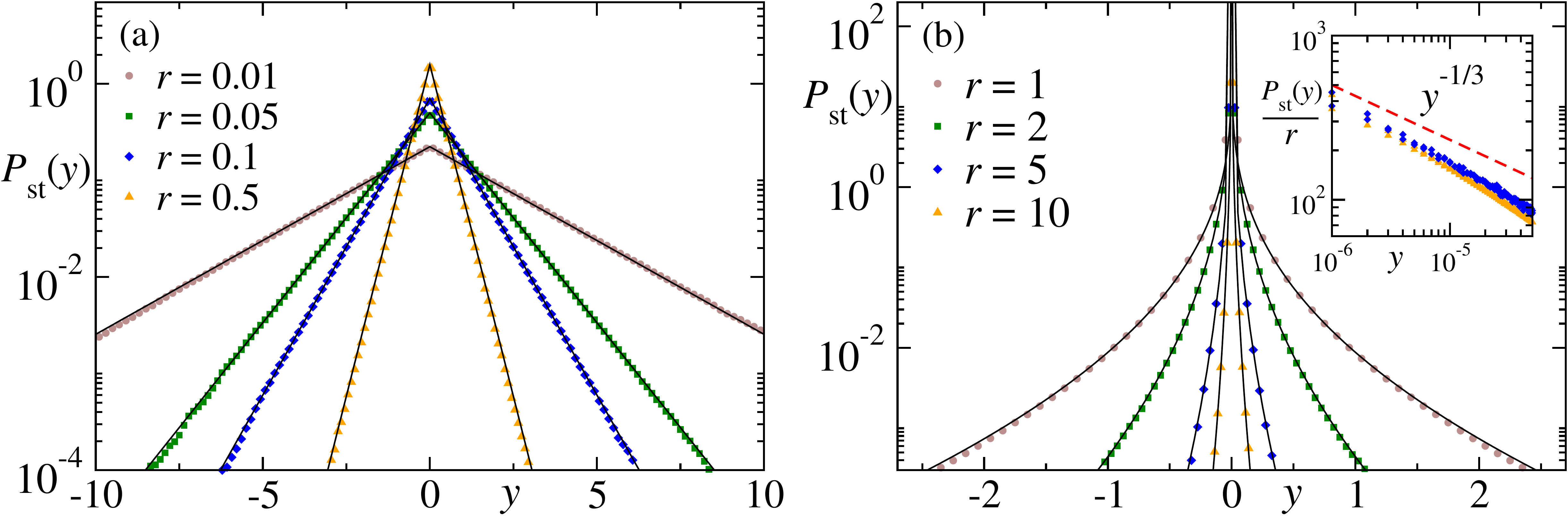} \hspace*{0.2cm}\includegraphics[width=5.7 cm]{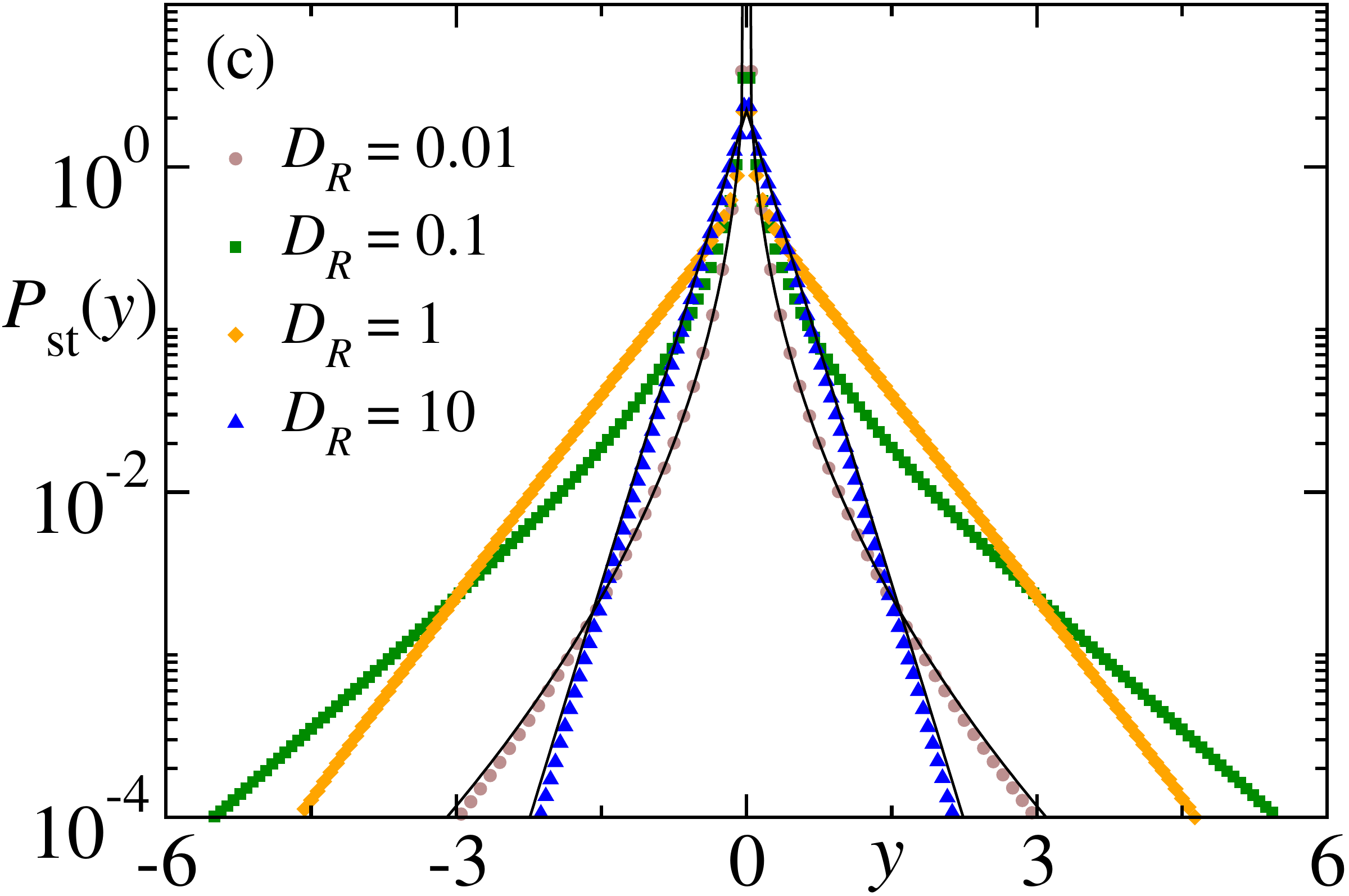} 
    \caption{Marginal stationary distribution $P_\text{st}(y)$ for position-orientation resetting: (a) Plot of $P_\text{st}(y)$ versus $y$ for different values of $r$ in the regime $r \ll \dr$ and a fixed $\dr=10.$ (b) Similar plot in the regime $r \gg \dr$ with  $\dr=0.01.$ The inset shows the algebraic divergence near the origin [see Eq.~\eqref{eq:Pst_pos_th_ysmall}].  Panel (c) shows the crossover between the two regimes for a fixed value of $r=1$. In all the plots the solid black lines indicate the analytical predictions [see Eq.~\eqref{eq:Py_rsmall} for (a) and Eq.~\eqref{eq:Py_rlarge} for (b)]. The symbols indicate the data obtained from numerical simulations. For all the plots $v_0=1$.}
    \label{fig:pdf_pos_th_reset_y}
\end{figure*}

\noindent {\bf Small resetting rate ($r \ll \dr$):} 
In this case, as before, we can use the late time expression for the ordinary active Brownian particle. In fact, at late times $t \gg \dr^{-1},$ free ABP loses the anisotropy, and the marginal $y$-distribution becomes same as the marginal $x$-distribution. Thus, the typical $y$-fluctuations are also Gaussian, and we can use Eq.~\eqref{eq:P0xs} for $P_0(y,t).$ Obviously, this leads to the same stationary exponential distribution, 
\bea   
P_\text{st}(y) = \frac 1{v_0} \sqrt{\frac {r\dr}2} \exp{\left[-\sqrt{2r \dr} \frac{|y|}{v_0} \right]}. \label{eq:Py_rsmall}
\eea   
This analytical prediction is verified in Fig.~\ref{fig:pdf_pos_th_reset_y}(a)  which shows a plot of the predicted $P_\text{st}(y)$ versus $y$ for different values of $r$ in the regime $r \ll \dr$ along with the same obtained from numerical simulations.  \\

\noindent {\bf Large resetting rate ($r \gg \dr$):}
Following the same argument as in the previous section, we expect that in this case, the stationary distribution can be determined from  the short-time behaviour of $P_0(y,s).$ Note that, because of the strong anisotropic nature of the free ABP at short-times, $P_0(y,s)$ is very different than $P_0(x,s)$ used in the previous section. In fact, it has been shown~\cite{ABP2018} that, at short-times $s \ll \dr^{-1},$ the $y$-dynamics of the ABP can be mapped to a Random Acceleration Process and $P_0(y,s)$ has a Gaussian form with variance $\frac 23 v_0^2 \dr s^3$ [see Appendix \ref{app:freeABP_pos} for more details]. Then the stationary $y$-distribution in the presence of resetting is given by,
\bea       
P_\text{st}(y) = \frac{\sqrt{3}r}{2 v_0 \sqrt{\pi \dr}} \int_0^{\infty}\id s~ \frac{e^{-r s}}{s^{3/2}} \exp{\bigg[-\frac{3y^2}{4 v_0^2 \dr s^3}\bigg]}.~~
\eea
It is useful to use a change of variable $u=rs,$ which leads to,
\bea
P_\text{st}(y) = \frac{\sqrt{3}r^{3/2}}{2 v_0 \sqrt{\pi \dr}} \int_0^{\infty} \id u~ \frac{e^{-u}}{u^{3/2}}\exp{\bigg[-\frac{3 r^3 y^2}{4 v_0^2 \dr u^3}\bigg]}.~~
\eea 
Clearly, the stationary distribution is a function of the scaled variable $z=\frac{r^{3/2}y}{v_0\sqrt{\dr}}.$ In fact, this integral can be computed exactly using Mathematica and the stationary distribution can be expressed in a scaling form,
\bea
P_\text{st}(y) = \frac{2 \pi r^{3/2}}{3 v_0 \sqrt{3 \dr}} \cal F\bigg(\frac{r^{3/2}|y|}{v_0\sqrt{\dr}}\bigg),\label{eq:Py_rlarge}
\eea
where the scaling function,
\bea
\cal F(z) &=& \frac 3{\pi^2} \bigg[\text{ker}_{1/3}\left(2 \sqrt{\frac z 3}\right)^2 + \text{kei}_{1/3}\left(2 \sqrt{\frac z 3}\right)^2\bigg].\quad \label{eq:Fz}
\eea
Here $\text{kei}_\nu(w)$ and $\text{ker}_\nu(w)$ are Kelvin functions (see Eq. 10.61.2 in Ref.~\cite{dlmf}). It can be shown that the stationary distribution given by Eqs.~\eqref{eq:Py_rlarge} and \eqref{eq:Fz} is identical to Eq.~(19) of Ref.~\cite{Prashant} obtained in the context of  resetting of Random Acceleration Process. 


Figure~\ref{fig:pdf_pos_th_reset_y}(b) shows a plot of the predicted stationary distribution for different (large) values of $r$ along with the same measured from numerical simulations.



It is interesting to look at the asymptotic behaviour of this stationary distribution. The behaviour near the origin can be obtained using the series expansion of the Kelvin functions. The details are provided in the Appendix \ref{app:asymp_Psty}; here we just quote the final result.  As $ |y| \to 0,$ $P_\text{st}(y)$ shows  an algebraic divergence,   
\bea       
P_\text{st}(y) = \frac{2 \pi r}{(v_0^2 \dr)^{1/3} 3^{7/6} \Gamma(\frac 23)^2} |y|^{-1/3} + \cal O(1). \label{eq:Pst_pos_th_ysmall}
\eea 
The inset in Fig.~\ref{fig:pdf_pos_th_reset_y}(b) shows a log-log plot of $P_\text{st}(y)$ near the origin where this divergence is illustrated.

To understand the decay of the distribution for large $|y|,$ we use the asymptotic expansion of the Kelvin functions for large argument; see Appendix \ref{app:asymp_Psty} for the details.  This exercise leads to a compressed exponential form for large $z,$ 
\bea
P_\text{st}\left(z = \frac{r^{3/2}y}{v_0 \sqrt{\dr}}\right) \simeq \frac{3 \sqrt{3}}{4\pi \sqrt{z}}\exp{-\sqrt{\frac{8z}{3}}}. \label{eq:Pst_pos_th_largey}
\eea
Figure \ref{fig:pdf_pos_th_reset_y}(b) shows a plot of $P_\text{st}(y)$ versus $y$ for different values of $r \gg \dr$ obtained from numerical simulations along with the analytical predictions.    

    
To investigate the crossover between the limiting cases ($r \ll \dr$ and $r \gg \dr$), we use numerical simulations.  Figure \ref{fig:pdf_pos_th_reset_y}(c) shows a plot of $P_\text{st}(y)$ versus $y$ for different values of $\dr$ and fixed $r=1.$
As expected, the divergence near the origin disappears as $\dr$ is increased. Moreover,  we see that, with increasing $\dr$, the width of the distribution first increases, and then decreases again, consistent with Eq.~\eqref{eq:vary_pos_th}.

\section{ABP with position resetting}\label{sec:protocolII}

In this Section we focus on the behaviour of the ABP under resetting protocol II, \ie, the position-resetting. In this case, the particle position is reset to the origin $x=y=0$  with rate $r$, but the orientation is not affected by the resetting events. As before, we consider that the particle starts from the origin with $\theta=0$ at time $t=0.$ Hence, at any time $t,$ the $\theta$ distribution remains Gaussian with zero-mean and variance $2\dr t.$ 

Our objective is to find the position distribution $P(x,y,t)$.  We can derive a renewal equation for the same in the following way. Let us consider the evolution of the particle trajectory during the interval $[0,t].$ If there are no resetting events during this interval, the position evolves under ordinary active Brownian motion. For the trajectories with at least one resetting, let us consider that the time elapsed since the last resetting event is given by $s.$ Then, the position at time $t$ is dictated by the free ABP evolution during this interval $s,$ but starting from some arbitrary orientation $\theta_{t-s},$ which itself is dictated by the Brownian motion of $\theta.$ Then the position distribution is obtained by integrating over all possible values of $0 \le s \le t,$ and $\theta_{t-s} \in [-\infty, \infty].$ 
Combining all these contributions, we get the renewal equation, 
\bea
P(x,y,t) &=& e^{-r t} P_0(x,y,t) + r \int_0^t \id s~ e^{-r s} \times \cr
& & \int_{-\infty}^{\infty} \id \theta~ \mathbb{P}_0^\theta(x,y,s) \frac{e^{-\frac{\theta^2}{4\dr(t-s)}}}{\sqrt{4 \pi \dr(t-s)}},\label{eq:pos_res_renew}
\eea
where we have used the notation $\mathbb{P}_0^\theta(x,y,s)$ to denote the probability that the free ABP is at $(x,y)$ at time $s,$ starting from an initial orientation $\theta$ at $s=0.$ The structure of the above renewal equation is different than the same obtained for the position-orientation resetting [see Eq.~\eqref{eq:renewal_pos_th}], and the behaviour is also expected to be different. 


The renewal equations for marginal distribution can be obtained by integrating over either $x$ or $y.$ We will investigate the stationary marginal position distributions later in Sec.~\ref{sec:dist_protocolII}. In the following, we first look at the moments to get an idea about the nature of the motion. 

\subsection{Moments}\label{sec:moments_protocolII}

The time-evolution of the moments of the position can be obtained from Eq.~\eqref{eq:pos_res_renew} in a straightforward manner. Let us first look at the moments of the $x$-position. Multiplying Eq.~\eqref{eq:pos_res_renew} by $x^n$ and integrating over both $x$ and $y,$ we get a renewal-like equation for the $n-$th moment of the $x$-component of the position,
\bea
\la x^n(t) \ra &=& e^{-r t} \la x^n(t)\ra_0 \cr
&+& r \int_0^t \id s~ e^{-r s} \int \id \theta~ \la x^n(s)\ra_0^\theta \frac{e^{-\frac{\theta^2}{4\dr(t-s)}}}{\sqrt{4 \pi \dr(t-s)}}.\cr
&& \label{eq:xmoments_pos_reset}
\eea
Here $\la x^n(s)\ra_0^\theta$ denotes the corresponding $n^{th}$ moment for the free ABP, starting from the origin, but oriented along some arbitrary direction $\theta$ and   $\la x^n(t)\ra_0,$ as before, denotes the moment starting from $\theta=0.$  Similarly, we can also write an equivalent renewal equation for the $y$-moments. The free ABP moments appearing in Eq.~\eqref{eq:xmoments_pos_reset} can be calculated exactly, and Appendix \ref{app:free_abp} provides explicit form for $n=1$ and $2.$
We use these expressions to calculate the first two moments of $x$ and $y$ for this position-resetting protocol.

 \begin{figure}[t]
    \centering
    \includegraphics[width=8.8 cm]{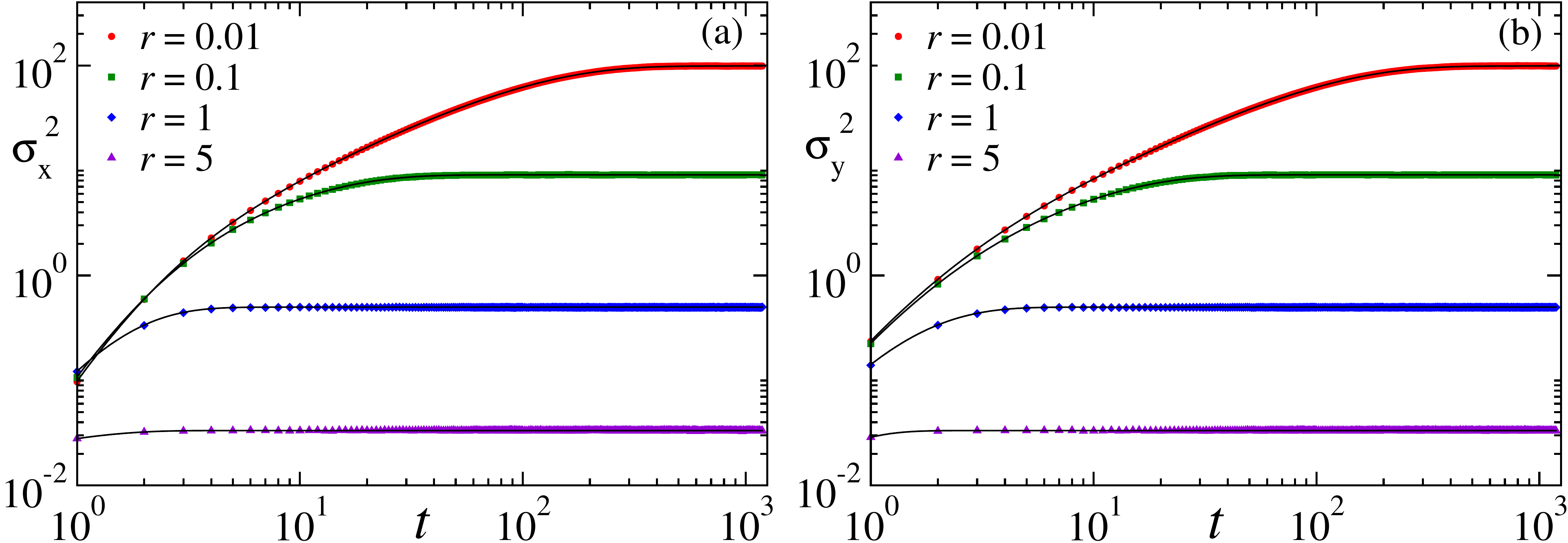}
    \caption{Position resetting: Variance of $x$-coordinate (a) and $y$-coordinate (b) as  functions of time $t$ for different values of $r$ and a fixed $\dr=1.$ The symbols correspond to the data from numerical simulations whereas the solid black lines indicate the analytical results [see Eqs.~\eqref{eq:pos_reset_varxt} and \eqref{eq:pos_reset_varyt}]. $v_0=1$ here. }
    \label{fig:Moments_pos_reset}
\end{figure} 

Using Eqs.~\eqref{eq:xt_yt_theta0_pos_reset} and \eqref{eq:xt_yt_0} in Eq.~\eqref{eq:xmoments_pos_reset}, we get the time-evolution of the average position,
\bea
\la x(t) \ra &=& \frac {v_0}{\dr-r} (e^{-r t}-e^{-\dr t}),\label{eq:pos_reset_xt_yt}
\eea
and $\la y(t) \ra = 0.$ Note that, for $r=\dr$ the above equation remains well defined, with $\la x(t)\ra = v_0 t e^{-\dr t}.$ Clearly, the average position approaches the origin in the stationary state $t \to \infty.$ At short times, \ie, for $t \ll \min(r^{-1},\dr^{-1}),$
\bea 
\la x(t) \ra = v_0t - v_0 (r+\dr) t^2 + O(t^3),
\eea
indicating that the resetting does not change the effective velocity to the leading order. 

  \begin{figure*}[t]
    \centering
    \includegraphics[width=11.83cm]{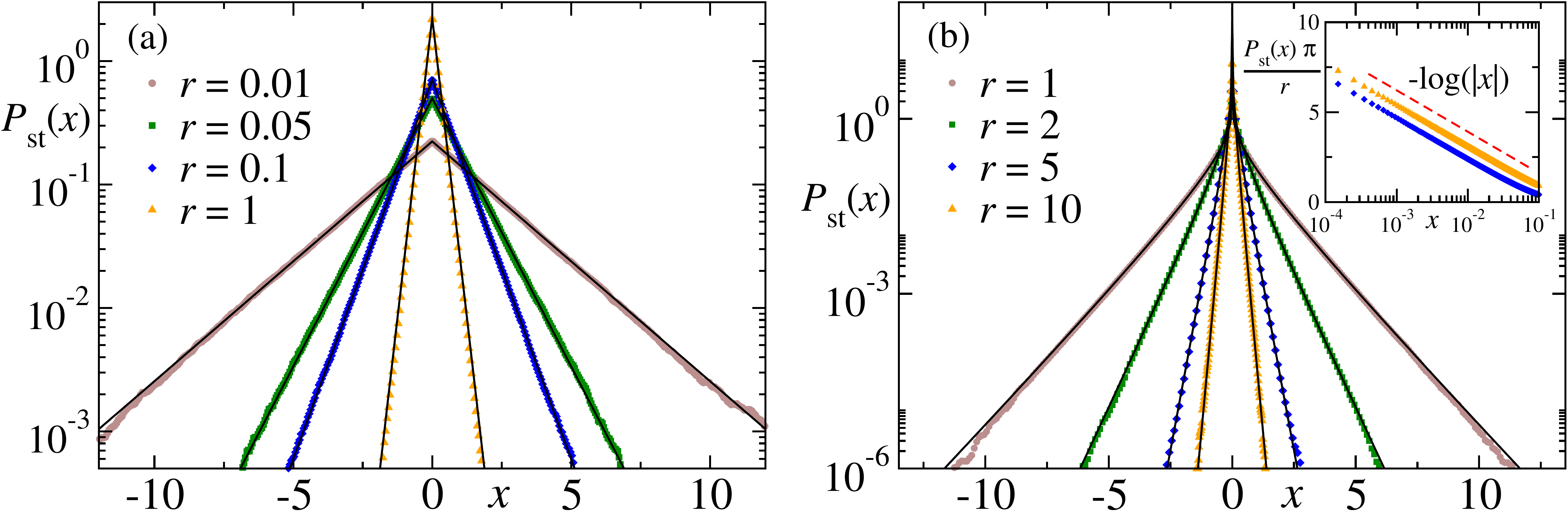} \hspace{0.25 cm} \includegraphics[width=5.65cm]{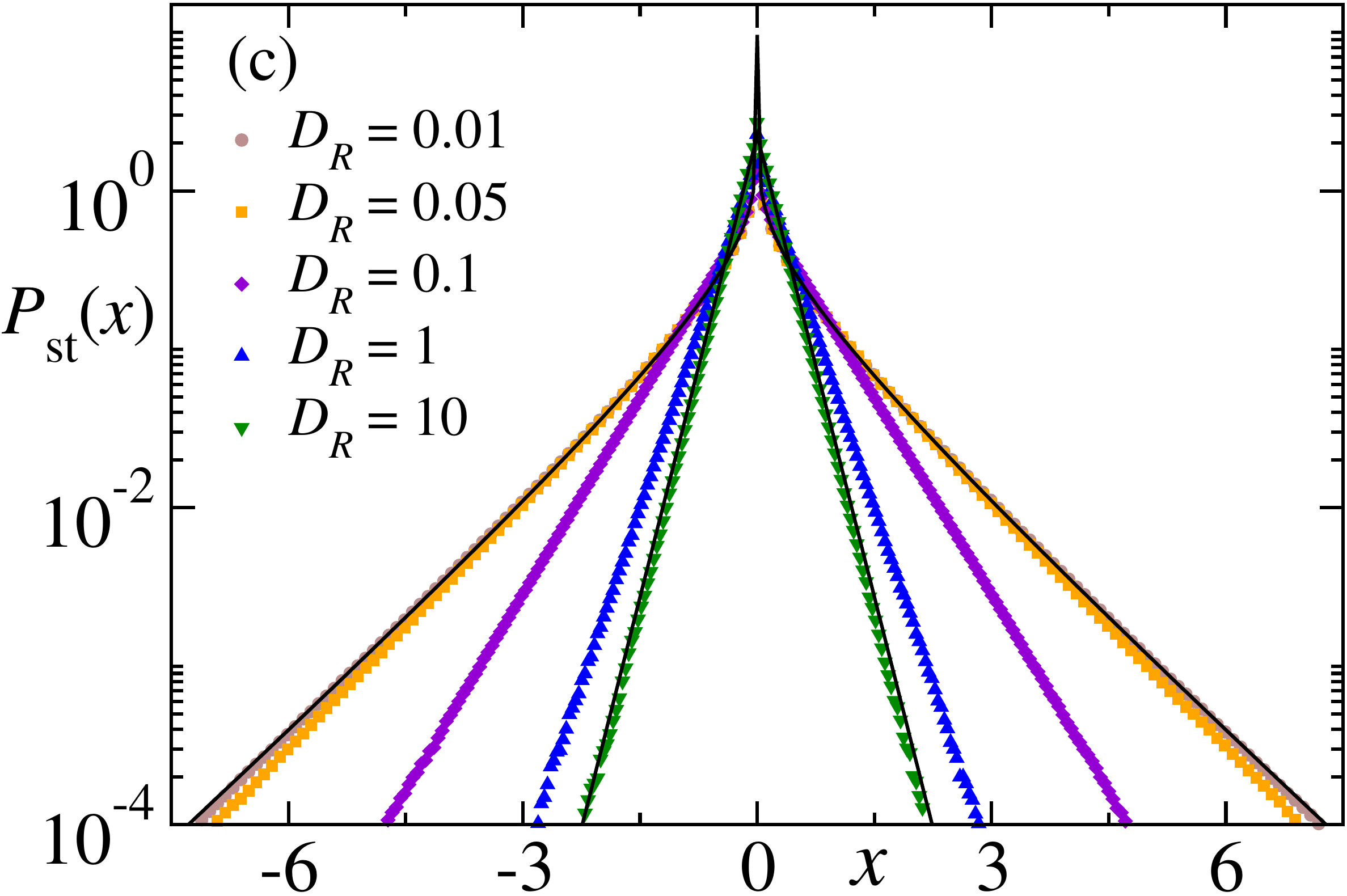}
    \caption{Position resetting: Stationary $x$-marginal probability distribution  (a) Plot of $P_\text{st}(x)$ versus $x$ for different values of $r$ in the regime $r \ll \dr$ with $\dr=10.$ (b) $P_\text{st}(x)$ versus $x$ in the regime $r \gg \dr$ with $\dr=0.01.$ Inset shows the logarithmic divergence near the origin [see Eq.~\eqref{eq:pos_log_div}]. (c) Plot of $P_\text{st}(x)$ covering both the limiting cases with $r=1$ and for different values of $\dr.$  In all the plots the numerical simulation results are indicated by symbols and the solid black lines indicate the analytical predictions; see Eq.~\eqref{eq:Pst_pos_large_D} for (a) and Eq.~\eqref{eq:Pstx_rlarge} for (b). Panel (c) shows that, for $r \gg \dr$ the distribution becomes independent of $\dr.$ $v_0=1$ here.  }
    \label{fig:px_pos_reset}
\end{figure*} 

The second moment of the $x$ and $y$-components can also be calculated exactly using Eq.~\eqref{eq:xmoments_pos_reset}. The explicit expressions are provided in Eqs.~\eqref{eq:pos_reset_x2t} and \eqref{eq:pos_reset_varyt} in the Appendix \ref{app:pos_mom}, here we quote the short-time and long-time behaviour of Mean squared displacements of x and y components. At short-times $t \ll \min(r^{-1},\dr^{-1}),$ we have,
\bea
\sigma_x^2(t) &=& \frac{v_0^2}{3}  r t^3+\frac{v_0^2}{12} (4 \dr^2-3 \dr r -4 r^2) t^4 + O(t^5),\cr
\sigma_y^2(t) &=& \frac{2 v_0^2}{3}   \dr t^3 -\frac{v_0^2}{6}  \dr(5 \dr+2  r) t^4  + O(t^5), \label{eq:varxy_short_pos}
\eea
indicating a superdiffusive behaviour. Moreover, even though both the variances show $t^3$ growth in this regime, the coefficients are different, which is a signature of the anisotropy present in the short-time regime. On the other hand, at late times $t \gg \max(r^{-1},\dr^{-1}),$, both $\sigma_x^2(t)$ and $\sigma_y^2(t)$ reach the same stationary value,
\bea
\sigma_x^2 = \sigma_y^2=  \frac{v_0^2}{r(\dr+r)}, \label{eq:varxy_long_pos}
\eea
indicating that the anisotropy disappears in the steady state. Figure \ref{fig:Moments_pos_reset}(a) and (b) show plots of $\sigma_x^2(t)$ and $\sigma_y^2(t)$ for different values of $r$ along with the same obtained from numerical simulations.  

In the next section we discuss the stationary probability distribution for this position-resetting protocol.

\subsection{Marginal position distribution}  \label{sec:dist_protocolII}
 
In the presence of position resetting only, the position distribution satisfies the renewal equation \eqref{eq:pos_res_renew}. As before, we focus on the stationary distribution, which is obtained by taking $t \to \infty$ limit. Clearly, the first term drops off in this limit. In the second term, the presence of the $e^{-rs}$  implies that the dominant contribution of the integrand comes from the regime $s < r^{-1}.$ For any finite $r,$ then, in the limit of large $t,$  $t-s \simeq t,$ and the Gaussian factor becomes flat. Now, since $\mathbb{P}_0^\theta(x,y,s)$ is a periodic function of $\theta,$ we can reduce the $\theta$-integral over one period, say, to the interval $[-\pi,\pi]$ where $\theta$ is distributed uniformly. The stationary distribution can then be expressed as,
\bea
P_\text{st}(x,y) = \frac{r}{2\pi} \int_0^\infty \id s ~e^{-r s} \int_{-\pi}^{\pi} \id \theta ~\mathbb{P}_0^\theta(x,y,s). \label{eq:Pxy_st_pos}
\eea
The $\theta$-integration makes the stationary distribution isotropic and it suffices to look at the marginal distribution along $x$-axis only. Integrating over $y,$ we get from Eq.~\eqref{eq:Pxy_st_pos},
\bea
P_\text{st}(x) = \frac{r}{2\pi} \int_0^\infty \id s ~e^{-r s} \int_{-\pi}^{\pi} \id \theta ~ \mathbb{P}_0^\theta(x,s). \label{eq:Px_st_pos}
\eea
We proceed as in the previous section, looking at the two limiting cases, namely, $r\ll \dr$ and $r \gg \dr.$ We also follow the same reasoning outlined in the previous section, and identify the region which contributes dominantly to the integral in \eqref{eq:Px_st_pos} in the two limiting cases. \\

\noindent {\bf Small resetting rate ($r \ll \dr$):} In this case, the dominant contribution to the integral \eqref{eq:Px_st_pos} comes from the long-time behaviour of free ABP distribution $\mathbb{P}_0^\theta(x,s).$ At long-times $s \gg \dr^{-1},$ the anisotropy disappears, and the distribution does not depend on the initial orientation $\theta.$ In fact, as mentioned in the previous section, to the leading order the long-time distribution is a Gaussian (see Appendix \ref{app:freeABP_pos}). Using this Gaussian form for $\mathbb{P}_0^\theta(x,s)$ in Eq.~\eqref{eq:Px_st_pos}, we get an exponential stationary distribution,
\bea
P_\text{st}(x) = \frac 1{v_0} \sqrt{\frac {r\dr}2} \exp{\left[-\sqrt{2r \dr} \frac{|x|}{v_0} \right]},\label{eq:Pst_pos_large_D}
\eea 
which is same as in the $r \ll \dr$ regime for the  position-orientation resetting case. 

Figure~\ref{fig:px_pos_reset}(a) compares the above prediction with the data from numerical simulations for a set of (small) values of $r$ and a fixed $\dr.$ An excellent match over a large range of $r$ illustrates the validity of Eq.~\eqref{eq:Pst_pos_large_D}, along with  the underlying assumptions. \\

\noindent {\bf Large resetting rate ($r \gg \dr$):} In this case, the stationary distribution is dominated by the contributions from the short-time trajectories of the free ABP, but starting from an arbitrary angle $\theta.$ As the behaviour of free ABP is ballistic at short-times $s \ll \dr^{-1}$, as a first approximation we can use (see  Appendix \ref{app:freeABP_pos}), 
\bea 
\mathbb{P}_0^\theta(x,s) \simeq  \delta (x - v_0 s \cos \theta). \label{eq:px0_th}
\eea
Using the above equation in \eqref{eq:Px_st_pos}, and performing the integrals (see Appendix \ref{app:dist_rlarge} for details), we get, 
\bea
P_\text{st}(x) = \frac r {\pi v_0} K_0\left(\frac{r |x|}{v_0}\right), \label{eq:Pstx_rlarge}
\eea
where $K_0(w)$ is the modified Bessel function of second kind \cite{dlmf}. Interestingly, within this approximation, the stationary distribution does not depend on the rotational diffusion constant $\dr$ at all  in this large $r$ limit. This is in contrast to the position-orientation resetting, where the limiting distribution depends on both $r$ and $\dr.$  Figure~\ref{fig:px_pos_reset}(b) shows a plot of $P_\text{st}(x)$ predicted in Eq.~\eqref{eq:Pstx_rlarge} for a set of (large) values of $r$ and a fixed $\dr$ along with the same obtained from numerical simulations; the excellent agreement confirms our analytical prediction. 
 
It is interesting to look at the asymptotic behaviour of the stationary distribution given in Eq.~\eqref{eq:Pstx_rlarge}. Expanding $K_0(w)$ near $w=0$, we find that, the distribution shows a logarithmic divergence near the origin, 
\bea
P_\text{st}(x) = - \frac r{\pi v_0} \log |x| + O(1).\label{eq:pos_log_div}
\eea
The inset in Fig.~\ref{fig:px_pos_reset}(b) illustrates this logarithmic divergence.  
On the other hand, for large $x \gg v_0/r$ the distribution falls off exponentially, 
\bea
P_\text{st}(x) \simeq  \sqrt{\frac{r}{2 v_0 |x|}} \exp{\bigg[- \frac {r|x|}{v_0}\bigg]}. \label{eq:pos_large_x}
\eea

It should be mentioned that we have restricted to the leading order approximate forms for the free ABP to calculate the position distribution in both the limiting scenarios. We can improve the range of validity (in $r$) of the analytical predictions by using next order corrections. However, in that case the integrals cannot be evaluated analytically and the  qualitative behaviour remains the same. Hence we skip this exercise here.  

 We use numerical simulations to investigate the crossover of the stationary distribution between the two limiting cases discussed above. Figure~\ref{fig:px_pos_reset}(c) shows a plot of $P_\text{st}(x)$ for a fixed $r$ and a range of values of $\dr;$ as $\dr$ is increased from the regime $\dr \ll r,$ the divergence near the origin disappears, and the distribution crosses over to the exponential behaviour. The width of the distribution also decreases continuously as $\dr$ is increased, as expected from Eq.~\eqref{eq:varxy_long_pos}.

\section{ABP with orientation resetting}\label{sec:protocolIII}
In this Section we consider the third resetting protocol where the orientation   $\theta$ resets to $\theta=0$ with rate $r$ while the position does not. In this case the position distribution does not satisfy any renewal equation directly but the $\theta$-distribution does. Let $\cal P(\theta, t| \theta',t')$ denote the probability that the orientation takes the value $\theta$ at time $t$ given that it was $\theta'$ at an earlier time $t'.$ $\cal P(\theta, t| \theta',t')$ satisfies a renewal equation \cite{Brownian},
\bea
\cal P(\theta,t|\theta',t') &=& e^{-r (t-t')} \cal P_0(\theta,t|\theta',t') \cr
&+& r \int_0^{(t-t')} \id s~ e^{-r s} ~\cal P_0(\theta,s|0,0),\label{eq:th_renewal}
\eea
where $\cal P_0 (\theta,t|\theta',t')$ denotes the propagator for the standard Brownian motion, given by Eq.~\eqref{eq:Brownian_prop}.  At long-times, the orientation reaches a stationary state with an exponential distribution although the position does not.  As before, we look at the moments of $x$ and $y$ components, and the corresponding  marginal position distributions. 

\subsection{Moments}   

The Langevin equations \eqref{eq:langevin} can be formally integrated to write,
\bea
x(t) &=& v_0 \int_0^t \id s \cos \theta(s), \cr
y(t) &=& v_0 \int_0^t \id s \sin \theta(s), \label{eq:xt_yt}
\eea 
where we used the initial condition $x(0)=y(0)=0.$ 
To calculate the position moments we need to know the mean and the auto-correlations of $\cos \theta$ and $\sin \theta$ under resetting which can be calculated using the propagator \eqref{eq:th_renewal}. The details of this calculation is provided in the Appendix \ref{app:th_reset_mom}, here we just quote the results.   
As in all the previous cases, $\la y(t) \ra$ vanishes at all times due to symmetry. Along $x$-axis, however, the average displacement is given by,
\bea
\la x(t) \ra &=&  \frac {v_0}{r+\dr} \bigg[rt+ \frac{\dr}{r+\dr}\bigg(1-e^{-(r+\dr)t}\bigg)\bigg],~~~ \\ \label{eq:xtav_proIII}
&=& \left \{ \begin{split}
v_0 t \;\; \; & \text{for}~ t \ll (r+\dr)^{-1},\cr
\frac{v_0r t}{\dr+r} &  \text{for}~ t \gg (r+\dr)^{-1}. \\
\end{split}
\right.
\eea

\begin{figure}[t]
    \centering
    \includegraphics[width=8.8 cm]{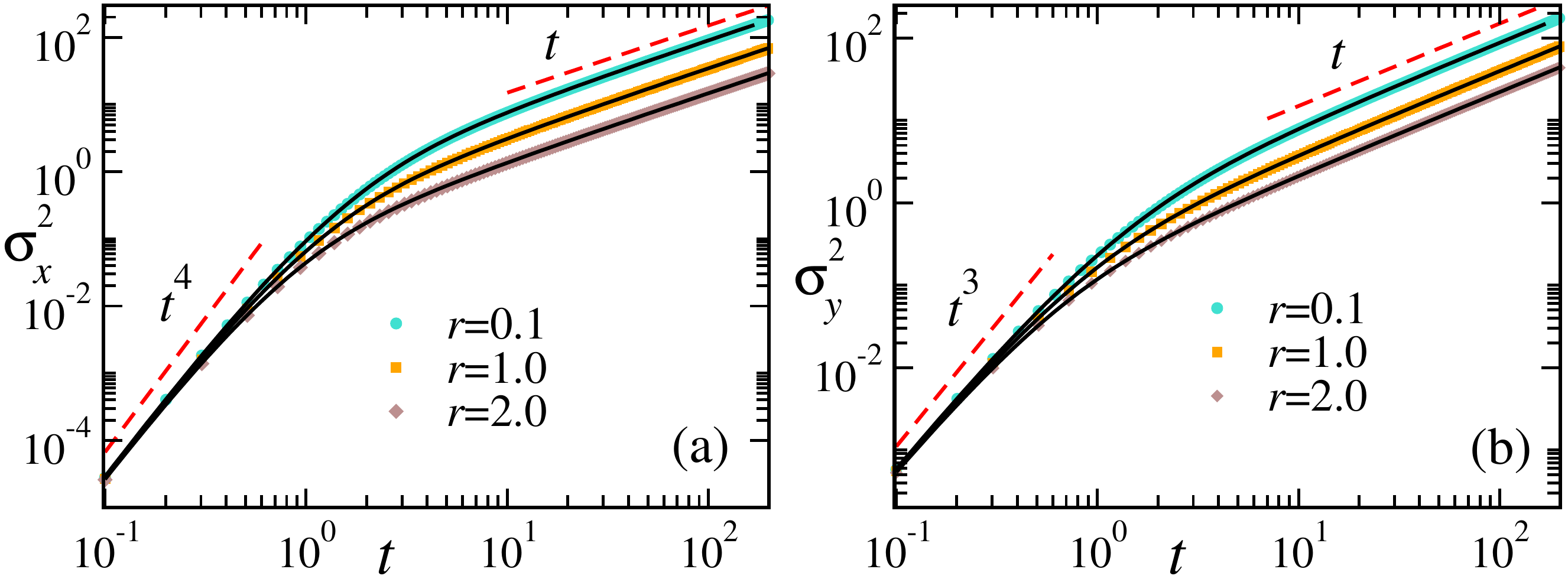}
    \caption{Orientation resetting: Plot of $\sigma_x^2$ (a) and $\sigma_y^2$ (b)  versus time $t$ for different values of $r$ with $\dr=1$ and $v_0=1$ The solid lines correspond to the analytical predictions Eqs.~\eqref{eq:varx_th_reset} and \eqref{eq:y2t_th_reset}. The red dashed lines indicate the predicted behaviour in the short-time and long-time regimes. }
    \label{fig:Moments_th_reset}
\end{figure} 

Clearly, the $x$-motion is ballistic at short-times with the velocity $v_0,$ which is reminiscent of the free ABP. Unlike the previous cases considered here, the effective velocity remains non-zero at late times, however, its value changes to $v_\text{eff}= \frac{v_0 r}{\dr +r}$ due to the presence of the resetting. 

\begin{figure}[t]
    \centering
    \includegraphics[width=8.8 cm]{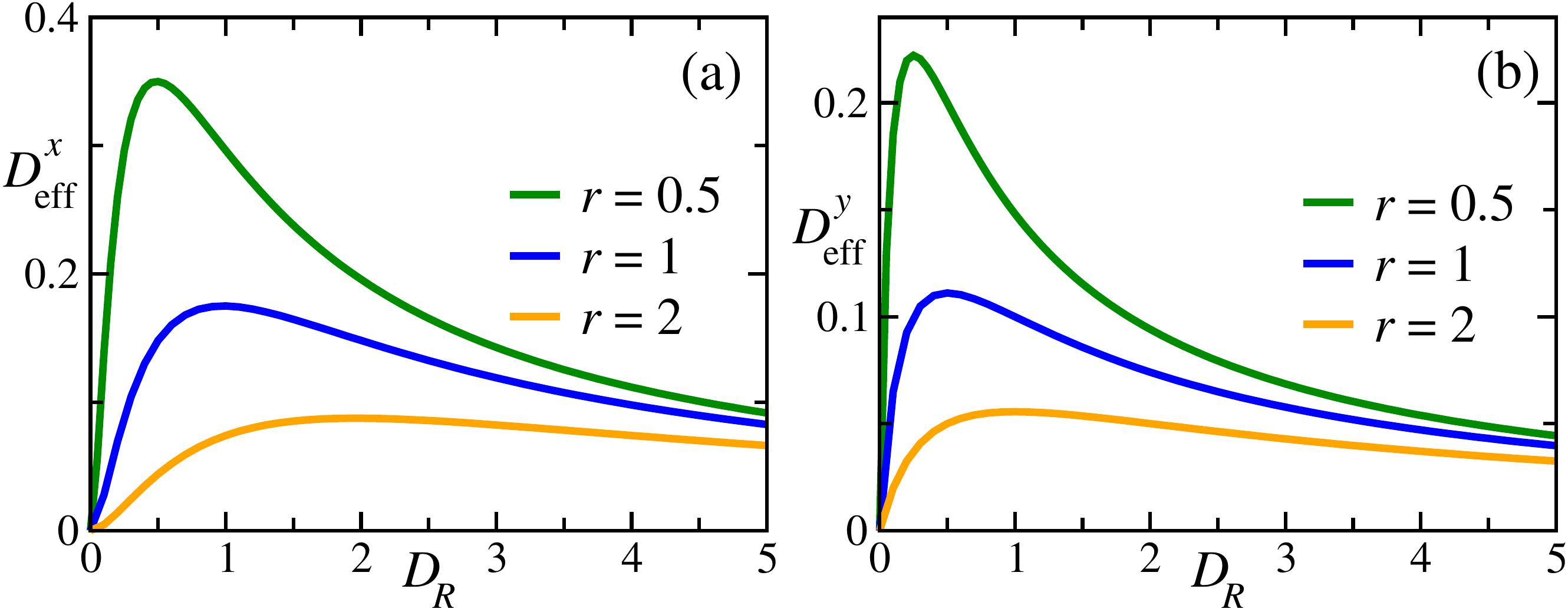}
    \caption{Orientation resetting: Plots of $D_\text{eff}^x$ (a) and $D_\text{eff}^y$ (b)  versus $\dr$ for different values of $r$.  See Eqs.~\eqref{eq:Deff_th} for the analytical expressions. We have taken $v_0=1.$ }
    \label{fig:Deff_th_reset}
\end{figure}

To understand the fluctuations around the mean position, we also look at the mean-squared displacement. The exact and long expressions for $\la x^2(t) \ra$ and  $\la y^2(t) \ra$ are provided in Eqs.~\eqref{eq:x2t_th_reset} and \eqref{eq:y2t_th_reset} respectively in Appendix \ref{app:th_reset_mom}.  
These analytical predictions are compared with numerical simulation results in Fig.~\ref{fig:Moments_th_reset} for different values of $r$ and a fixed $\dr.$  As in the previous cases, we see that both the $x$ and $y$-variances show a crossover from a superdiffusive to a diffusive behaviour as time $t$ increases.  To understand the nature of these crossovers, we look at the  short-time and long-time behaviours of the mean-square displacements. At very short-times, \ie, for $t \ll (r+\dr)^{-1},$ we have,
\bea 
\sigma_x^2(t)&=& \frac{v_0^2}{3}\dr^2 t^4-\frac{ v_0^2}{30} \dr^2 (14\dr-5r)t^5+\cal O(t^5), \cr 
\sigma_y^2(t) &=& \frac{2 v_0^2}{3} \dr  t^3-\frac{v_0^2 }{6}\dr(5 \dr+2r) t^4+\cal O(t^5).~~~\label{eq:varxy_short_th}
\eea  
To the leading order, this behaviour is same as that of free ABP  with strong anisotropy between $x$ and $y$ motions~\citep{ABP2018}. The effect of resetting appears at higher orders, and it introduces an additional anisotropy. 
This is expected, as the resetting configuration $\theta=0$ is also strongly anisotropic. The effect of this anisotropy sustains at late-times also -- even though both $x$ and $y$ motions become diffusive, \ie, 
\bea
\lim_{t \to \infty}  \sigma_x^2 \simeq   2 D_\text{eff}^x ~ t, \quad \lim_{t \to \infty}  \sigma_y^2 \simeq   2 D_\text{eff}^y ~ t, \label{eq:varxy_longt_th}
\eea
the effective diffusion constants remain very different,
 \bea
D_\text{eff}^x &= & \frac{v_0^2 \dr^2 (2 \dr +5r)}{(4\dr+r)(\dr+r)^3},  \cr    
 D_\text{eff}^y & =& \frac{2v_0^2 \dr}{(\dr+r)(4\dr+r)}. \label{eq:Deff_th}
 \eea
Figure \ref{fig:Deff_th_reset} shows plots of $D_\text{eff}^x$ and $D_\text{eff}^y$ as functions of  $\dr,$ for a set of values of $r.$ It is interesting to note that these effective diffusion constants are non-monotonic in $\dr$ -- for a fixed $r,$ $D_\text{eff}^{x,y}$ reach their corresponding maximum values for some intermediate values of $\dr$ which increases as $r$ is increased. 

  
\subsection{Marginal position distributions}
To understand the behaviour of the position distribution, let us first look at a trajectory with $n$ resetting events during the interval $[0,t]$. Let us also assume that $t_i$ denotes the interval between the $i$ and $(i-1)$-th resetting event. At any time $t,$ the position $(x(t),y(t))$ can be expressed as a sum of position increments over the intervals $t_i,$ 
 \bea   
x(t) &=& \sum_{i=1}^{n+1}  x_0(t_i), \label{eq:x_sum} \\
y(t) &=& \sum_{i=1}^{n+1}  y_0(t_i). \label{eq:y_sum}
\eea 
Let us remember that, in between the resetting events the system evolves as an ordinary ABP and hence, the fluctuations of $x_0(t_i)$ and $y_0(t_i)$ follow the  distribution $P_0(x_i,y_i,t_i),$ where we have used the notation $x_i \equiv x_0(t_i)$ and $y_i \equiv y_0(t_i)$

As before, we focus on the marginal distributions of $x$ and $y$-components separately.  From Eq.~\eqref{eq:x_sum}, the $x$-distribution in the presence of orientation resetting can be formally written as,
\bea  
P(x,t) &=& \sum_{n=0}^\infty r^n e^{-r t} \int \prod_{i=1}^{n+1} \id t_i \id x_i ~P_0(x_i,t_i) \cr
&& \times \delta\bigg(x-\sum_{i=1}^{n+1} x_i\bigg)\, \delta\bigg(t-\sum_{i=1}^{n+1} t_i \bigg), \label{eq:Pxt_series}
\eea   
where $P_0(x_i,t_i)$ denotes the probability that, in the absence  of resetting, the ABP has a displacement $x_i$ during the time-interval $t_i,$ starting from $\theta=0.$  
The $y$-marginal distribution also has a similar form,
\bea  
P(y,t) &=& \sum_{n=0}^\infty r^n e^{-r t} \int \prod_{i=1}^{n+1} \id t_i \id y_i P_0(y_i,t_i) \cr
&& \times \delta\bigg(y-\sum_{i=1}^{n+1} y_i \bigg)\,\delta \bigg(t-\sum_{i=1}^{n+1} t_i \bigg),  \label{eq:Pyt_series}
\eea 
where $P_0(y_i,t_i)$ denotes the probability that the $y$-component of the position of the free ABP has a displacement $y_i$ during the interval $t_i,$ staring from $\theta=0.$ Let us note that $P_0(x_i,t_i)$ and $P_0(y_i,t_i)$ have different functional forms, in particular for small $t_i,$ even though we have used the same letter for notational simplicity. 

It is hard to compute the marginal distributions from the above equations exactly, as explicit form for the position distributions in the absence of resetting  are not known. However, as we will see below, we can still understand the different behaviours in the short and long-time regimes.

\begin{figure} 
\includegraphics[width=6.5 cm]{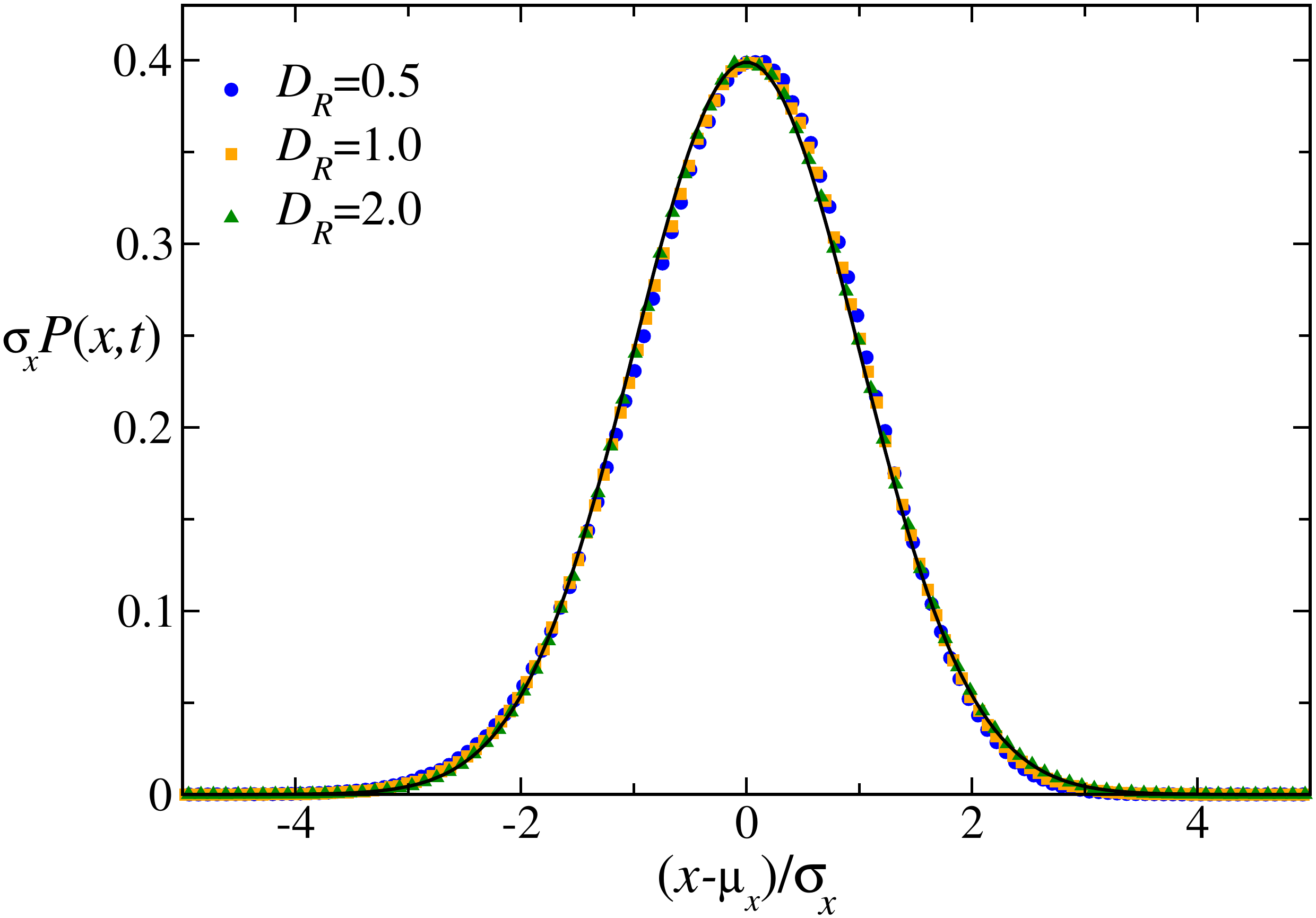}
\caption{Orientation resetting: Plot of the scaled marginal $x$-distribution at a long-time $t=500$ for $r=1$ and different values of $\dr.$ The solid black line shows the standard normal distribution. $v_0=1$ here. \label{fig:px_longt}}
\end{figure}

Let us first focus at the long-time regime. As indicated by the moments, we expect a diffusive motion for both $x$ and $y$-components in this regime. For simplicity, let us first consider the $x$-component. From Eq.~\eqref{eq:x_sum}, we see that the net displacement along $x$-direction is given by a sum of $n+1$ random variables, namely, the displacements during the intervals $t_i.$ Since, after each reset, the orientation $\theta$ is brought back to its initial value, and the time-evolution starts afresh, the  variables $x_0(t_i)$ are independent and identically distributed (of course, the duration $t_i$ are different). Even though  the distribution of $x_i$ is not known explicitly, its moments are all finite. Over a large time interval $t,$ the number $n$ of the resetting events is typically large, with $\la n \ra = rt.$ For $t \gg r^{-1}$ then $x(t)$ is a sum of a large number $n$ of independent and identically distributed random variables. From central limit theorem, we can then expect that $x(t)$ has a Gaussian distribution,
\bea
P(x,t) = \frac 1{\sqrt{2 \pi \sigma_x^2(t)}} \exp{\left[-\frac{(x-\mu_x(t))^2}{2 \sigma_x(t)^2}\right]}, \label{eq:px_gaussian}
\eea  
where $\mu_x(t)=\la x(t) \ra$ and $\sigma_x(t)^2$ are the mean and variance given by Eqs.~\eqref{eq:xtav_proIII} and \eqref{eq:varxy_longt_th} (with large $t$). Note that this prediction is independent of the value of $r;$ for each $r,$ there exists some $t \gg r^{-1}$ above which we expect a Gaussian distribution, albeit with different $r$-dependent means and variances.  Figure \ref{fig:px_longt} shows a plot of $\sigma_x(t) P(x,t)$ vs $(x- \mu_x(t))/\sigma_x(t)$ for $r=1,t=500$ and different values of $\dr;$ a perfect collapse verifies the prediction.   

\begin{figure}
\includegraphics[width=6.5 cm]{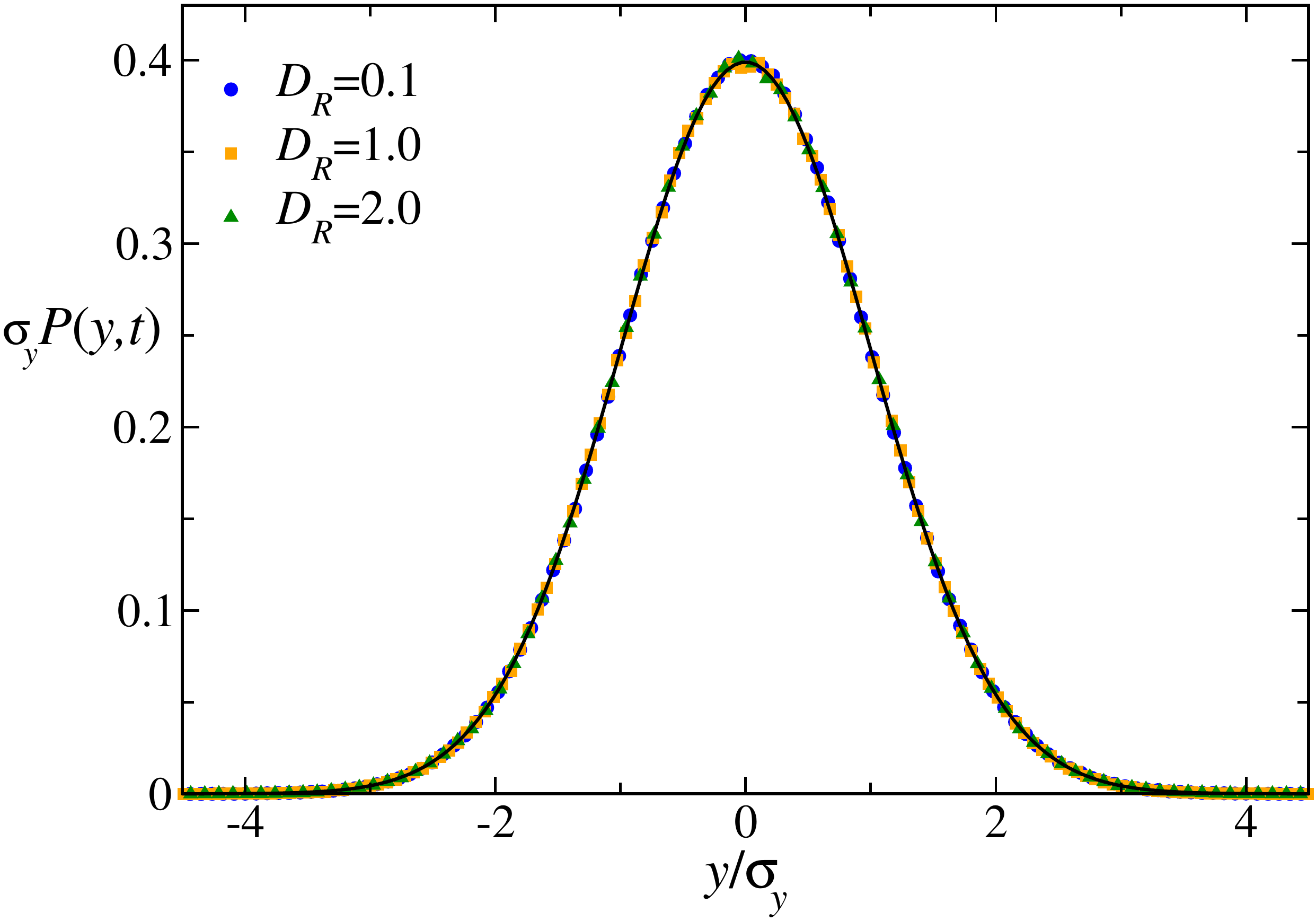}
\caption{Orientation resetting: Plot of the scaled marginal $y$-distribution at a long-time $t=500$ for $r=1$ and different values of $\dr$ obtained from numerical simulations. The solid black line shows the standard normal distribution.$v_0=1$ here.  \label{fig:py_longt}}
\end{figure}

The same argument can be applied to $y(t),$ from Eq.~\eqref{eq:y_sum}, and we expect,
\bea 
P(y,t) = \frac 1{\sqrt{2 \pi \sigma_y^2(t)}} \exp{\left[-\frac{y^2}{2 \sigma_y(t)^2}\right]}, \label{eq:py_gaussian}
\eea  
where $\sigma_y^2(t)$ is the large $t$-behaviour obtained from Eq.~\eqref{eq:varxy_longt_th}. We also observe a perfect collapse for $P(y,t)$, as depicted in Figure \ref{fig:py_longt} which verifies our prediction.

\begin{figure}
\includegraphics[width=6.5 cm]{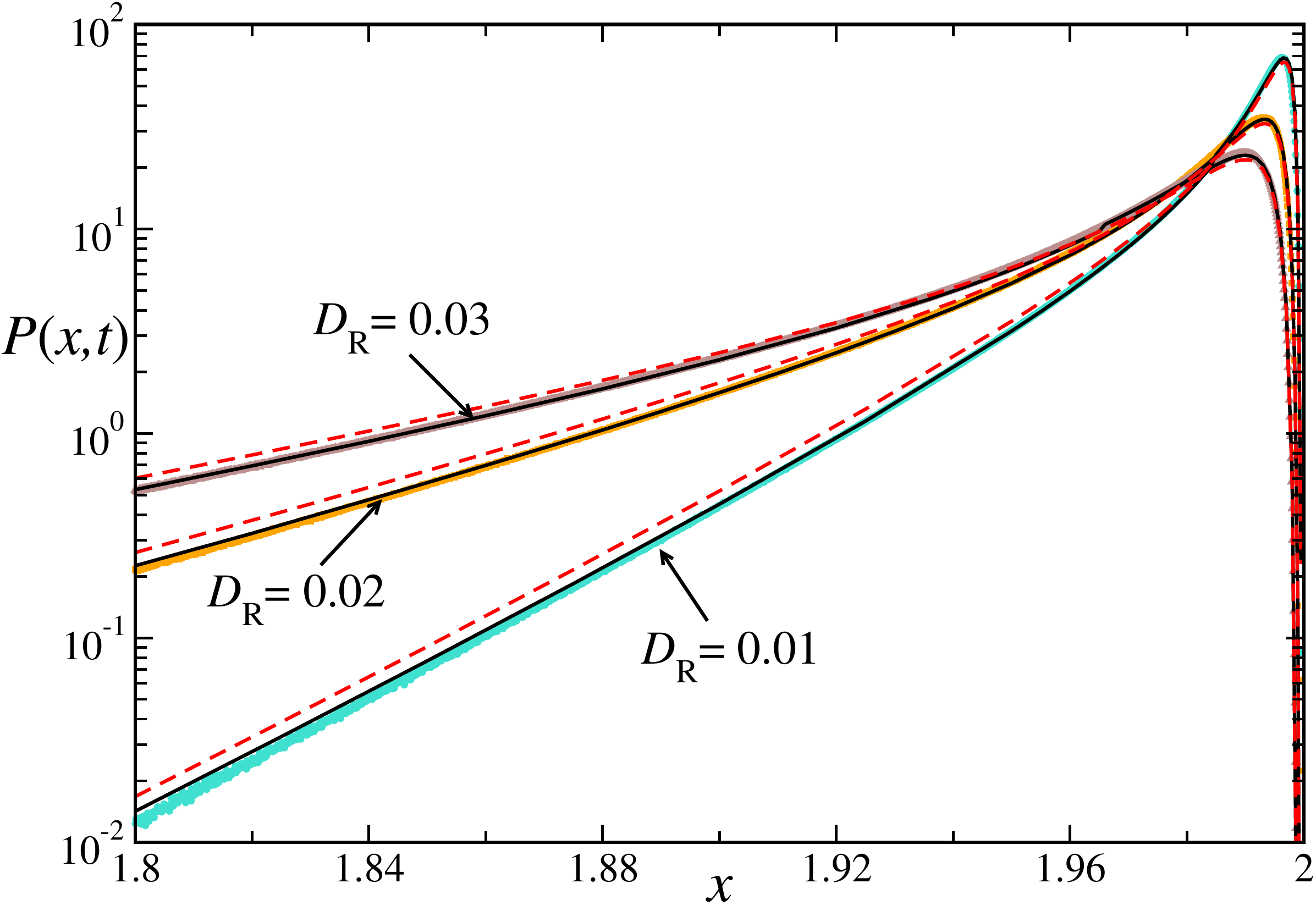}
\caption{Orientation resetting: Plot of the marginal $x$-distribution at a short-time $t=2$ for $r=0.1$ and different values of $\dr.$ We have taken $v_0=1.$  \label{fig:px_short_t} } 
\end{figure}

In the  short-time regime, the average number of resetting events is small and we can expect small $r$ contributions to dominate. From Eq.~\eqref{eq:Pxt_series}, one can adopt a perturbative approach, that is, for small $r$ we compute the distribution at short-times. 
In fact, to obtain the leading order correction introduced by the resetting, we truncate the sum  after $n=1,$ which is equivalent to keeping linear order in $r$ (apart from the $e^{-rt}$ factor). 
We then get,  
\bea  
P(x,t) &=&  e^{-rt}[P_0(x,t)+ r P_1(x,t)+ \cal O(r^2) ].    
\eea    
Here $P_0(x,t)$ is the short-time marginal $x$-distribution for the active Brownian particle without resetting given in Eq.~\eqref{eq:P0xt_shortt} and $P_1(x,t)$ is the leading order correction due to resetting,
\bea
P_1(x,t) =  \int_0^t \id t_1 \int_{a(t_1)}^ {b(t_1)}  \id x_1 ~P_0(x_1,t_1) 
P_0(x-x_1,t-t_1).~~\label{eq:P1x}
\eea
The limits on the $x_1$-integral are determined  from the condition that $P_0(x,t)$ is non-zero only in the region $ -t \le x \le t$ and are given by, 
\bea
a(t_1) &=& \text{max}(-t_1, x-t+t_1),\cr
b(t_1) &= & \text{min}(t_1, x+t-t_1).
\eea
Using Eq.~\eqref{eq:P0xt_shortt} the integrals in Eq.~\eqref{eq:P1x} can be evaluated numerically with arbitrary accuracy. The resulting $P(x,t),$ which is expected to be valid in the regime $t \ll \dr^{-1},$ is plotted in Fig.~\ref{fig:px_short_t}  for different (small) values of $\dr$ and a fixed (small) values of $r$ and $t$ along with the same obtained from numerical simulations. The analytical prediction matches well with the results from simulation indicating that the perturbative approach works fairly well in this regime. The position distribution appears similar in shape to that in the absence of resetting, with a peak near $x=v_0t.$ However, quantitatively they are different, as can be seen from the plot --- we have included the corresponding curves for $r=0$  as dashed lines for easy comparison. Clearly, the effect of resetting becomes more pronounced away from the peak.  

We follow the same perturbative procedure to compute the $y$-marginal distribution also. From Eq.~\eqref{eq:Pyt_series}, we write, to the leading order in $r,$    
\bea
P(y,t) &=& e^{-rt}[P_0(y,t) + r P_1(y,t) + \cal O(r^2)],\label{eq:Py_pert}
\eea 
with,
\bea
P_1(y,t) =  \int_0^t \id t_1 \int_{\tilde a(t_1)}^ {\tilde b(t_1)}  \id y_1 ~P_0(y_1,t_1) 
P_0(y-y_1,t-t_1).~~\label{eq:P1y}
\eea
As before, the integration limits are obtained from the condition that $ -t_1 \le y_1 \le t_1$ and $ t_1 -t \le y - y_1 \le t-t_1,$
\bea
\tilde a(t_1) &=& \text{max}(-t_1, y-t+t_1)\cr
\tilde b(t_1) &= & \text{min}(t_1, y+t-t_1).
\eea

\begin{figure}
\includegraphics[width=6.5 cm]{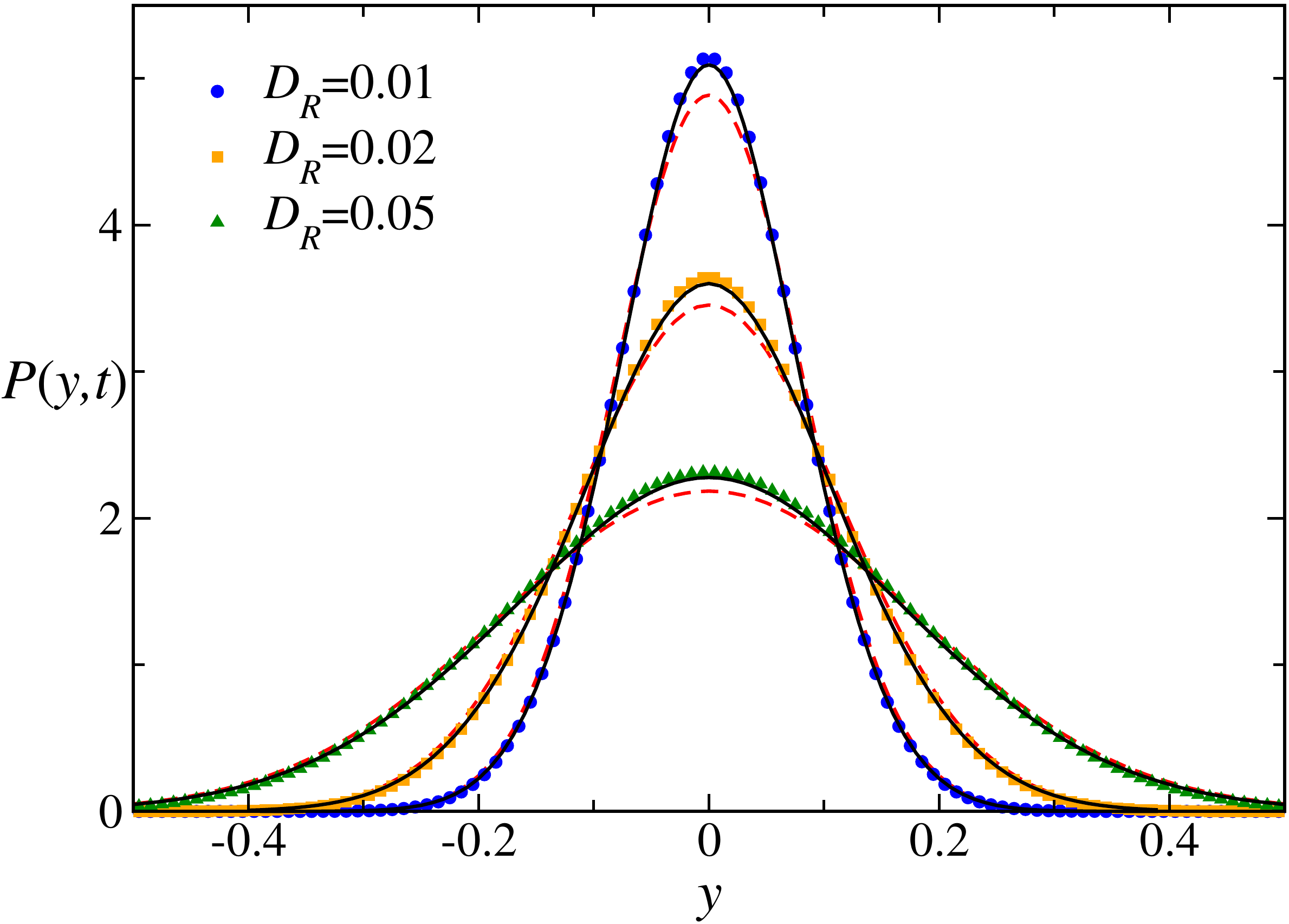}
\caption{Orientation resetting: Plot of the marginal $y$-distribution at a short-time $t=1$ for $r=0.1$ and different values of $\dr.$ The solid black lines correspond to the analytical prediction Eq.~\eqref{eq:Py_pert} and the dashed red lines correspond to Gaussian ($r=0$) curves. We have taken $v_0=1.$ } \label{fig:py_short_t} 
\end{figure}

We obtain $P_1(y,t)$ by numerically evaluating the integral in Eq.~\eqref{eq:P1y}. As before, we restrict ourselves in the regime $t \ll \dr^{-1},$ so that the short-time expression of $P_0(y,t)$ [see Eq.~\eqref{eq:P0yt_shortt}] is applicable. 
The resulting marginal distribution $P(y,t)$ is plotted in Fig.~\ref{fig:py_short_t} for a set of values of $\dr$ with a fixed (small) $r=0.1$ and $t=1$ along with the same obtained from numerical simulations. The distribution has a single peak at the origin, similar to the $r=0$ case (indicated by dashed lines) in shape. However, the correction due to resetting makes it non-Gaussian, the difference with $r=0$ case is clearly visible near the peaks.

\section{Conclusions} \label{sec:concl}

We study the position distribution of an active Brownian particle in 2D under stochastic resetting. An ABP is characterized by its position as well as an internal orientation. We show that depending on whether the resetting protocol affects the position degrees of freedom or the orientational degree, the ABP shows a wide range of rich behaviour. In particular, we study three different resetting protocols, namely, resetting both position and orientation to their initial value, resetting only the position, and resetting only the orientation.  We find that in the first two cases the position reaches stationary states. We show that the interplay between the time-scales due to resetting and the rotational diffusion leads to a set of different regimes --  depending on whether the resetting rate $r$ is smaller or larger than the rotational diffusion constant $\dr,$ the stationary distributions take very different shape.  Using renewal approach, we compute exactly the marginal distributions of the $x$ and $y$-components in the limiting cases $r \ll \dr$ and $r \gg \dr.$

In the first case, \ie, when both the position and orientation are reset to their initial value, we find that, for small resetting rates $ r \ll \dr,$ the marginal distributions $P_\text{st}(x)$ and $P_\text{st}(y)$ are exponential in nature, with the same decay exponent. On the other hand,  for $r \gg \dr,$ the position distribution becomes strongly anisotropic. The marginal $x$-distribution is non-zero only for $x>0$ in this case, with an exponential decay at the tail, and approaching a finite value near the origin $ x \to 0^+.$ The $y$-distribution, which is symmetric, shows a very different behavior, with an algebraic divergence near the origin ($|y| \to 0$) and a compressed exponential decay at the tails. 

For the position-resetting case, the position distribution is isotropic for all values of $r$ and $\dr.$ For $r \ll \dr$ the distribution turns out to be exponential in nature. For large values of $r \gg \dr,$ the position distribution shows a logarithmic divergence near the origin, while decaying exponentially at the tails.

In the third case, \ie, when the resetting protocol affects only the orientation of the ABP, the position of the particle does not reach a stationary state, but continues to increase along the $x$-direction with an effective velocity. However, the nature of the motion changes from ballistic, at short-times, to diffusive, at late times ($t \gg (r+\dr)^{-1}$). We show that, at late times, the typical position fluctuations around the mean are characterized by Gaussian distributions for both $x$ and $y$ components, albeit with different effective diffusion constants. At short-times, the position distribution remains strongly non-Gaussian, which we characterize using a perturbative approach for small resetting rates.

The resetting of a particle in position space can be thought of as the effect of switching on and off an external trap, in the spirit of Ref. \cite{trap_reset, Schehrreset2020}. On the other hand, the orientation resetting can be envisaged as the effect of an external magnetic field on magnetic active particles \cite{mag1, mag2, mag3, mag4}, which is switched on at random times. In general, it is interesting to study what happens when the position and orientation resetting can occur independently of each other. Another obvious open question is how the persistence properties of the ABP are affected in the presence of such resetting mechanisms. It would be also interesting to see how introduction of resetting mechanism affects other active particle models and if a general picture emerges. The behaviour of active particles under different resetting protocols present another intriguing set of questions. 

\acknowledgements 
V. K. acknowledges  support from Raman Research Institute where he worked as a visiting student and where this work was carried out. O. S. acknowledges the Inspire grant from DST, India. U. B. acknowledges support from Science and Engineering Research Board, India under Ramanujan Fellowship (Grant No. SB/S2/RJN-077/2018).

\appendix

\section{Brief review of active Brownian motion in 2D}  \label{app:free_abp}

For the sake of completeness we provide a brief review of the free ABP dynamics in this Appendix. In the absence of resetting the position and orientation of the ABP evolves following the Langevin equation \eqref{eq:langevin}.  We assume that at time $t=0$ the particle starts from the origin $x=y=0,$ oriented along some arbitrary direction $\theta=\theta_0.$ As the orientation evolves following an ordinary Brownian motion the probability that the orientation is $\theta$ at time $t,$ given that it was $\theta'$ at some earlier time $t'$ is, 
\bea 
{\cal P}_0(\theta,t| \theta',t') = \frac 1 {\sqrt{4 \pi \dr (t-t')}} \exp{\bigg[-\frac{(\theta - \theta')^2}{4 \dr (t - t')}\bigg ]}.~~~ \label{eq:Brownian_prop}
\eea
In the following we quote the results for the moments and distribution of the position components $x$ and $y$ of the ABP.

\subsection{Moments of the position components} \label{app:moments_freeABP}

The moments of the position coordinate $x,y$ can be obtained in a straightforward manner \cite{ABP2018} using the Brownian propagator for the orientation $\theta$ given in Eq.~\eqref{eq:Brownian_prop}.  Here we compute the explicit expressions for the first two moments for arbitrary values of the initial orientation  $\theta_0.$
Integrating the Langevin equation \eqref{eq:langevin} and taking average over all possible trajectories, we have,
\bea
\la x(t) \ra^{\theta_0}_0 &=& v_0 \int_0^t \id s~ \la \cos \theta(s) \ra^{\theta_0}_0, \cr
\la y(t) \ra^{\theta_0}_0 &=& v_0 \int_0^t \id s~ \la \sin \theta(s) \ra^{\theta_0}_0,\label{eq:ABP_xt_yt_int}
\eea
where we have used the superscript $\theta_0$ to denote the initial orientation.
From Eq.~\eqref{eq:Brownian_prop} we have,
\bea 
\la \cos \theta(s) \ra^{\theta_0}_0&=& \int \id \theta~ \cos \theta(s)~ \frac{e^{-\frac{(\theta-\theta_0)^2}{4\dr s}}}{\sqrt{4 \pi \dr s}}\cr
&=&  \cos \theta_0 ~ e^{-\dr s},\label{eq:cos_thetas_ABP}
\eea
and similarly,
\bea
\la \sin \theta(s) \ra^{\theta_0}_0 =   \sin \theta_0 ~ e^{-\dr s}. \label{eq:sin_thetas_ABP}
\eea
The average positions can be computed using the above equations in Eqs.~\eqref{eq:ABP_xt_yt_int},
\bea
\la x(t) \ra^{\theta_0}_0 &=& \frac{v_0}{\dr} \cos \theta_0 \big(1-e^{-\dr t}\big),\cr
\la y(t) \ra^{\theta_0}_0 &=& \frac{v_0}{\dr} \sin \theta_0 \big(1-e^{-\dr t}\big).\label{eq:xt_yt_theta0_pos_reset}
\eea
Equations \eqref{eq:xt_yt_theta0_pos_reset} are used in Sec.~\ref{sec:moments_protocolII} to compute the first position moments in the presence of  the position resetting.

For computing the moments in the presence of the  position-orientation resetting, we need the ABP moments for $\theta_0=0.$ In this case  Eqs.~\eqref{eq:xt_yt_theta0_pos_reset} reduce to,
\bea
\la x(t) \ra_0 &=& \frac{v_0}{\dr}  \big(1-e^{-\dr t}\big), \cr
\la y(t) \ra_0 &=& 0.\label{eq:xt_yt_0}
\eea
which have been used in Sec.~\ref{sec:moments_protocolI} to compute the average positions.

Next, we look at the second moments. From Eq.~\eqref{eq:langevin}, we can write,
\bea
\la x^2(t) \ra^{\theta_0}_0 &=& 2~v_0^2\int_0^t \id s~\int_0^s \id s'~ \la \cos \theta (s) \cos \theta (s') \ra^{\theta_0}_0, \cr
\la y^2(t) \ra^{\theta_0}_0 &=& 2~v_0^2\int_0^t \id s~\int_0^s \id s'~ \la \sin \theta (s) \sin \theta (s') \ra^{\theta_0}_0.\cr
&&\label{eq:x2t_y2t_theta0}
\eea
The two-point correlations $\la \cos \theta (s) \cos \theta (s') \ra^{\theta_0}$ and $\la \sin \theta (s) \sin \theta (s') \ra^{\theta_0}$  can be calculated exactly using Eq.~\eqref{eq:Brownian_prop};  for $s> s'$ we get,

\bea
\la \cos \theta (s) \cos \theta (s') \ra^{\theta_0}_0 &=& \frac{1}{2} e^{-\dr (s-s')}\bigg[ 1+e^{-4\dr s'} \cos 2 \theta_0\bigg], \cr
\la \sin \theta (s) \sin \theta (s') \ra^{\theta_0}_0 &=& \frac{1}{2} e^{-\dr (s-s')}\bigg[1 -e^{-4\dr s'} \cos 2 \theta_0\bigg]. \cr
&&\label{eq:cos_sin_av}
\eea 
Now, substituting Eq.~\eqref{eq:cos_sin_av} in Eq.~\eqref{eq:x2t_y2t_theta0} and evaluating the integrals, we get,
\bea
\la x^2(t) \ra^{\theta_0}_0 &=& \frac{v_0^2}{12 \dr^2}\bigg[12(\dr t +e^{-\dr t}-1)\cr
&& + (3+e^{-4\dr t}-4 e^{-\dr t})\cos 2 \theta_0\bigg],\cr
\la y^2(t) \ra^{\theta_0}_0 &=& \frac{v_0^2}{12 \dr^2}\bigg[12(\dr t +e^{-\dr t}-1)\cr
&& -(3+e^{-4\dr t}-4 e^{-\dr t})\cos 2 \theta_0\bigg].\label{eq:x2t_y2t_ABP}
\eea
These expressions have been used in Sec.~\ref{sec:moments_protocolII} to compute the variances in the presence of position resetting.

Once again, for calculating the variances in the position-orientation case we need the expressions for  $\theta_0=0$, 
\bea
\la x^2(t)\ra_0 &=& \frac{v_0^2 t}{\dr} + \frac{v_0^2}{12 \dr^2} \left[e^{-4 \dr t} +8  e^{- \dr t}-9 \right], \cr
\la y^2(t)\ra_0 &=& \frac{v_0^2 t}{\dr} - \frac{v_0^2}{12 \dr^2} \left[e^{-4 \dr t} -16  e^{- \dr t}+15 \right].\cr
&& \label{eq:ABP_x2t_y2t}
\eea
which were obtained in \cite{ABP2018}. The above results are used in Appendix.~\ref{app:pos-th_rABP} to obtain Eqs.~\eqref{eq:x2t_rABP_pos_th} and \eqref{eq:varyt_pos_th}. 
%
It is also straightforward to calculate the third moment of $x(t)$ using Eq.~\eqref{eq:Brownian_prop}. For $\theta_0=0$ it turns out to be,
\bea
\la x^3(t)\ra_0 &=& \frac{1}{240 \dr^3} \left[ 5 e^{-\dr t} (120 \dr t+169)-4 e^{-4 \dr t}\right. \cr
&&\ \left. -e^{-9 \dr t} +720 \dr t -1340 \right].\label{eq:x3t_ABP}
\eea
This expression is used in Eq.~\eqref{eq:skew_lt} to compute the skewness; see also Appendix ~\ref{app:pos-th_rABP}.

%

\subsection{Position Distribution} \label{app:freeABP_pos}


In the absence of resetting, the position distribution  of the ABP is  given by $P_0(x,y,t) = \int \id \theta ~\cal P_0 (x,y,\theta,t)$ where $\cal P_0 (x,y,\theta,t)$ denotes the probability that the ABP has the position $(x,y)$ and orientation $\theta$ at time $t.$ $\cal P_0 (x,y,\theta,t)$ evolves according to the Fokker-Planck equation,
\bea
\frac {\partial \cal P_0} {\partial t}  = - v_0 \bigg[\cos \theta \frac {\partial \cal P_0 } {\partial x} + \sin \theta \frac {\partial \cal P_0 } {\partial y} \bigg] + \dr \frac {\partial^2 \cal P_0 } {\partial \theta^2}. 
\eea
Formally, the above equation can be solved using Fourier transformation with respect to position coordinates and the Fourier transform of $\cal P_0 (x,y,\theta,t)$  can be expressed in terms of an infinite series of Matthieu functions \cite{Franosch2018}. 
Unfortunately the Fourier transform cannot be inverted analytically, and no closed form expression for the position distribution is available. However, marginal position distributions for the $x$ and $y$  components, starting from  $\theta=0,$   in short-time and long-time regimes are known separately. For the sake of completeness, we quote these expressions here. 

In the short-time regime $(t \ll \dr^{-1})$, the marginal $x$-distribution can be expressed in a scaling form,
\bea
P_0(x,t)= \frac 1{v_0 \dr t^2} f_x\left(\frac{v_0 t -x}{v_0 \dr t^2}\right), \label{eq:P0xt_shortt}
\eea
where the scaling function is given by,
\bea
f_x(u)= \frac 1{2 \sqrt{\pi u^3}} \sum_{k=0}^\infty (-1)^k \frac {(4 k+1)}{2^{2k}} \left({2k \atop k}\right) e^{-\frac{(4 k+1)^2}{8 u}}. ~~~~~
\eea
The $y$-marginal distribution, on the other hand, has a Gaussian form in this short-time regime,
\bea
P_0(y,t)  = \frac {\sqrt 3}{2 v_0 \sqrt{\pi \dr t^3}} \exp {\bigg[- \frac {3y^2}{4 v_0^2 \dr t^3}\bigg]}. \label{eq:P0yt_shortt}
\eea
At late times $t \gg \dr^{-1}$ the anisotropy goes away, and it has been shown in Ref.~\cite{ABP2019} that in this regime both $x$ and $y$ marginal distribution admits a large deviation form, which is quoted in Eq.~\eqref{eq:large_dev_x}. \\


\noindent {\bf Initial orientation $\theta_0 \ne 0$:}  Next we look at marginal position distribution starting from any arbitrary $\theta_0 \ne 0.$ In this case,
we can substitute $\theta(t)= \theta_0 + \phi(t)$ in Eq.~\eqref{eq:langevin} 
where $\phi(t)$ undergoes a standard Brownian motion with $\phi(0)=0.$ 

At short-times $\phi(t) \sim \sqrt{t}$ is small, and to the leading order we can approximate $\sin \phi(t) \simeq \phi(t)$ and $\cos \phi(t) \simeq 1.$ In this regime, the Langevin equations \eqref{eq:langevin} reduce to,
\bea
\dot x(t) &\simeq & v_0[\cos \theta_0 - \phi(t) \sin \theta_0 ], \cr
\dot y(t) &\simeq & v_0 [\sin \theta_0 + \phi(t) \cos \theta_0].
\eea
Clearly, for non-zero $\theta_0$ both $x$ and $y$-components have systematic drifts.
To a first approximation, the position distribution can then be written as,
\bea
\mathbb{P}_0^{\theta_0}(x,y,t) = \delta(x- v_0 t \cos \theta_0)\delta(y- v_0 t \sin \theta_0), \label{eq:Pxy_th0}
\eea
where we have used the superscript $\theta_0$ to denote the initial orientation. 
The above expression, when integrated over $y,$ gives the $x$-marginal distribution quoted in Eq.~\eqref{eq:px0_th}. 
Note that here the fluctuation of the orientation is completely neglected. A better approximation is, of course, when the effect of $\dr$ is included, in which case the marginal distributions would be Gaussian. However, as shown in the Sec.~\ref{sec:dist_protocolII}, Eq.~\eqref{eq:Pxy_th0} suffices for computing the stationary distribution in the $r \gg \dr$ limit for the position resetting.

In the long-time limit $t \gg \dr^{-1}$ , on the other hand, the position distribution does not depend on the initial value of orientation and we expect the typical fluctuations to be Gaussian in nature, as given by Eq.~\eqref{eq:P0xs},
\bea 
\mathbb{P}_0^{\theta_0}(x,t) = \sqrt{\frac{\dr}{2 \pi v_0^2 t}}\, \exp{\bigg[- \frac{\dr x^2}{2 v_0^2 t}\bigg]}. 
\eea


\section{Exact computation of moments for position-orientation resetting} \label{app:pos-th_rABP}

In this Appendix we present the exact analytical expressions for the higher moments of the $x$ and $y$ components of position in the presence of position-orientation resetting. We can calculate the second moment of $x(t)$ from Eq.~\eqref{eq:xn_pos_th} as,
\bea
\la x^2(t) \ra &=& \frac{2 v_0^2(2\dr +r)}{r(\dr+r)(4\dr+r)} \n \\[0.25em]
&+& \frac{v_0^2 }{3\dr}e^{-rt} \left[\frac{e^{-4\dr t}}{4\dr+r} +\frac{2e^{-\dr t}}{\dr+r}- \frac 3r\right]. ~~\label{eq:x2t_rABP_pos_th}
\eea
Using Eqs.~\eqref{eq:xt_pos_th} and \eqref{eq:x2t_rABP_pos_th} we  obtain the variance $\sigma_x^2= \la x^2(t) \ra - \la x(t)\ra^2$,
\bea
\sigma_x^2 &=& \frac{v_0^2(4\dr^2+2r\dr +r^2)}{r(4\dr+r)(\dr+r)^2} + v_0^2 e^{-rt}\Bigg[\frac{e^{-4\dr t}}{3\dr (4\dr+r)} \cr
&-&\frac{e^{-(2\dr+r)t}}{(\dr+r)^2}+ \frac{2(4\dr+r)e^{-\dr t}}{3\dr (\dr+r)^2} - \frac 1{r\dr}\Bigg].\; \label{eq:varxt_pos_th}
\eea
The short-time and long-time limiting behaviour obtained from the above equation are quoted in the main text.

Similarly, we also calculate the variance $\sigma_y^2$, which is nothing but the second moment for the $y$-component. Using Eq.~\eqref{eq:ABP_x2t_y2t} in the renewal equation, we get,
\bea
\sigma_y^2 &=&\la y^2(t) \ra = \frac{4 v_0^2 \dr}{r(\dr+r)(4\dr+r)} \n \\[0.25em]
&-& \frac{v_0^2 }{3\dr}e^{-rt} \left[\frac{e^{-4\dr t}}{4\dr+r} -\frac{4e^{-\dr t}}{\dr+r}+ \frac 3r\right].~~\label{eq:varyt_pos_th}
\eea

To calculate the skewness of $P(x,t)$ we need the third moment. Using the expression of $ \la x^3(t)\ra_0$ given by Eq.~\eqref{eq:x3t_ABP} along with  Eq.~\eqref{eq:xn_pos_th}, we get, 
\bea
\la x^3(t) \ra &=& \frac{v_0^3}{240 \dr^2} \Bigg[\frac{1440 \dr^2 (3\dr+r)(6\dr+r)}{r(\dr+r)^2(4\dr+r)(9\dr+r)}\cr
&+& \frac{5 e^{-(\dr+r)t}}{(\dr+r)^2} \bigg (269 \dr+ 149 r +120 \dr (\dr+r)t\bigg) \cr
&-& e^{-r t} \bigg( \frac{720}{r}+\frac{16 e^{-4 \dr t}}{4 \dr+r}+\frac{9 e^{-9 \dr t}}{9 \dr+r}\bigg) \Bigg].\label{eq:x3t_pos_th}
\eea
The exact time-dependent expression for skewness $\gamma$ can be obtained using Eqs.~\eqref{eq:xt_pos_th},~\eqref{eq:varxt_pos_th} and ~\eqref{eq:x3t_pos_th}; we omit the rather long expression and quote the  stationary value $\gamma_\text{st}$ in Eq.~\eqref{eq:skew_limit} obtained by taking the limit $t \to \infty.$

\section{Asymptotic behaviour of $P_\text{st}(y)$ for position-orientation resetting}\label{app:asymp_Psty}

To find the behaviour of $P_\text{st}(y)$  for small and large values of $y,$ we use the asymptotic expansion of the Kelvin functions appearing in Eq.~\eqref{eq:Fz}. From the series expansion near $w=0,$ we have,
\bea
\text{ker}_{1/3}(w) &=& \frac{\Gamma(1/3)}{2^{7/6}} w^{-1/3} + O(w^{1/3}),\cr
\text{kei}_{1/3}(w) &=& - \frac{\Gamma(1/3)}{2^{7/6}} w^{-1/3} + O(w^{1/3}).
\eea
Inserting the above expressions in Eq.~\eqref{eq:Fz} along with Eq.~\eqref{eq:Py_rlarge} we get an algebraic divergence of  $P_\text{st}(y)$ near $y=0$ 
which is quoted in Eq.~\eqref{eq:Pst_pos_th_ysmall}.

On the other hand, for large values of the argument $w,$ we have (see Sec.~10.67 in Ref.~\cite{dlmf}),
\bea
\text{ker}_{1/3}(w) &=& e^{- w/{\sqrt 2}} \sqrt{\frac{\pi}{2w}} \cos \left(\frac w{\sqrt 2} + \frac {7 \pi}{24}\right) + O\left(\frac 1{w^{3/2}}\right), \cr
\text{kei}_{1/3}(w) &=& e^{- w/{\sqrt 2}} \sqrt{\frac{\pi}{2w}} \sin \left(\frac w{\sqrt 2} + \frac {7 \pi}{24}\right) + O\left(\frac 1{w^{3/2}}\right). \n
\eea
Using the above expressions along with Eqs.~\eqref{eq:Fz} and \eqref{eq:Py_rlarge}, we get the large $z$-behaviour of the scaling function quoted in Eq.~\eqref{eq:Pst_pos_th_largey}.


%

\section{Exact computation of moments for position resetting} \label{app:pos_mom}

In this Appendix we provide the explicit expressions for the second moments of the position in presence of the resetting protocol II, \ie, for only position resetting. Using the renewal equation \eqref{eq:xmoments_pos_reset} for $n=2,$ along with  Eqs.~\eqref{eq:x2t_y2t_ABP} and \eqref{eq:ABP_x2t_y2t}, we get, 
\bea 
 \la x^2(t) \ra &=&\frac{v_0^2}{\dr} \bigg[\frac{2(\dr-r) e^{-(\dr+r)t}}{(3\dr-r)(\dr+r)} - \frac{2(2\dr-r) e^{-r t}}{r(4\dr-r)}    \cr
 &+& \frac {\dr e^{-4\dr t} }{(4\dr-r)(3\dr -r)}+ \frac{\dr}{r(\dr+r)}\bigg]. \quad \label{eq:pos_reset_x2t}
\eea 
The variance can be calculated using the above equation along with Eq.~\eqref{eq:pos_reset_xt_yt} and is given by, 
\bea 
 \sigma_x^2(t) &=& v_0^2 \bigg[\frac{2 e^{-(\dr+r)t}}{(3\dr-r)(\dr+r)}\bigg(\frac{4 \dr^2}{(\dr-r)^2}-\frac{r}{\dr}\bigg) \cr
 &-& \frac{2 e^{-r t}(2\dr-r)}{r \dr (4\dr-r)} -\frac{(e^{-2 r t}+e^{-2 \dr t})}{(\dr-r)^2} \cr
 &+&   \frac {e^{-4\dr t} }{(4\dr-r)(3\dr -r)}+\frac{1}{r(\dr+r)}  \bigg]. \label{eq:pos_reset_varxt}
 \eea
To get the behavior in the short-time regime, \ie, for  $t \ll \min(r^{-1}, \dr^{-1})$ we can use the Taylor series expansion around $t=0.$ The resulting expansion is quoted in Eq.~\eqref{eq:varxy_short_pos}.  On the other hand, in the $ t \to \infty$ limit the variance reaches the stationary value quoted in Eq.~\eqref{eq:varxy_long_pos}.

The variance of $y(t)$ also satisfies the renewal equation \eqref{eq:pos_res_renew},
\bea
\la y^2(t) \ra  &=& e^{-r t} \la y^2(t)\ra_0 + r \int_0^t \id s~ e^{-r s} ~\times \cr
&&  \int_{-\infty}^\infty \id \theta~ \la y^2(s)\ra_0^\theta \frac{e^{-\frac{\theta^2}{4\dr(t-s)}}}{\sqrt{4 \pi \dr(t-s)}},\label{eq:y2_renewal_proII}
\eea
where $\la y^2(t)\ra_0,$ the second moment of ABP starting with $\theta=0$ is given by Eq.~\eqref{eq:ABP_x2t_y2t} and $\la y^2(t)\ra_0^\theta,$ the second moment of ABP starting with arbitrary orientation $\theta$ is given in Eq.~\eqref{eq:x2t_y2t_ABP}. Using these expressions in Eq.~\eqref{eq:y2_renewal_proII} we have,
\bea
  \la y^2(t) \ra &=&  \frac{v_0^2}{r(\dr+r)} +  v_0^2 \bigg[\frac{4 e^{-(\dr+r)t}}{(3\dr-r)(\dr+r)} \cr
   &-& \frac{4 e^{-r t}}{r(4\dr-r)}  - \frac {e^{-4\dr t} }{(4\dr-r)(3\dr -r)}\bigg].~~~\label{eq:pos_reset_varyt}
 \eea
The short-time behaviour is quoted in Eq.~\eqref{eq:varxy_short_pos} in the main text.

\section{Position resetting: marginal distribution for $r \gg \dr$}\label{app:dist_rlarge}

In this Appendix we provide the details of the calculation leading to Eq.~\eqref{eq:Pstx_rlarge}. Substituting $\mathbb{P}_0^\theta(x,s)$ from Eq.~\eqref{eq:px0_th} in Eq.~\eqref{eq:Px_st_pos}, we get the stationary distribution, 
\bea
P_\text{st}(x) &=& \frac r{2\pi} \int_0^\infty \id s~ e^{-r s} \int_{-\pi}^\pi \id \theta ~\delta(x - v_0 s \cos \theta)\cr
&=& \frac r{v_0 \pi } \int_0^\infty \id s~ e^{-r s}  \int_{0}^\pi \frac   {\id \theta}{|\cos \theta|}\delta\bigg(\frac x{v_0 \cos \theta} - s \bigg). \cr
&& \label{eq:px_pos_int}
\eea
Here, in the second step,  we have used the fact that $\cos \theta$ is an even function of $\theta.$ 
Now, for $x>0,$ the $\delta$-function contributes only when $\cos \theta >0,$ \ie, 
$0 \le \theta \le \frac \pi 2.$ Thus, evaluating the $s$-integral, we have, for $x>0,$
\bea
 P_\text{st}(x) &=& \frac r{v_0 \pi } \int_{0}^{\pi/2} \frac{\id \theta}{|\cos \theta|} \exp{\bigg[-\frac {rx}{v_0 \cos \theta}\bigg]}  \cr
 &=& \frac r{v_0 \pi } K_0 \left(\frac {rx}{v_0}\right). \label{eq:px_positive}
\eea
Here $K_0(z)$ is the modified Bessel function of the second kind. For $x<0,$ on the other hand, the $s$-integral in Eq.~\eqref{eq:px_pos_int} is non-zero only when $ \pi/2 \le \theta \le \pi.$ In this case we have,
\bea
 P_\text{st}(x) &=& \frac r{v_0 \pi } \int_{\pi/2}^{\pi} \frac{\id \theta}{|\cos \theta|} \exp{\bigg[-\frac {rx}{v_0 \cos \theta}\bigg]}  \cr
 &=& \frac r{v_0 \pi } K_0 \left(-\frac {rx}{v_0}\right). \label{eq:px_neg}
\eea
Combining Eqs.~\eqref{eq:px_positive} and \eqref{eq:px_neg} we get the complete marginal distribution quoted in Eq.~\eqref{eq:Pstx_rlarge}.

\section{Exact computation of the moments for the orientation resetting}\label{app:th_reset_mom}

To compute the moments of the position coordinates in the presence of the orientation resetting we  start from Eq.~\eqref{eq:xt_yt}. Taking statistical average over all possible trajectories of $\theta$, we get,
\bea
\la x(t) \ra &=& v_0 \int_0^t \id s~ \la \cos \theta(s) \ra, \cr
\la y(t) \ra &=& v_0  \int_0^t \id s~ \la \sin \theta(s) \ra. \label{eq:xt_yt_th}
\eea
The averages appearing on the right hand side can be computed using the renewal equation \eqref{eq:th_renewal} for $\cal P(\theta,t).$ We have, 
\bea
\la \cos \theta(s)\ra &=& \int_{- \infty}^\infty \id \theta \cos \theta ~ \cal P(\theta, s) \cr
&=& \frac{\dr}{\dr+r} e^{-(\dr+r)s}+\frac{r}{\dr+r}, \n
\eea
and $\la \sin \theta(s) \ra=0.$  Using the above expression in Eq.~\eqref{eq:xt_yt_th} we get the mean $x$-position which is quoted in Eq.~\eqref{eq:xtav_proIII}. Obviously, $\la y(t) \ra=0.$ \\

\noindent {\bf Variance of $x(t)$ and $y(t)$:} From Eq.~\eqref{eq:xt_yt} we have,
\bea
\la x^2(t) \ra &=& v_0^2 \int_0^t \id s \int_0^t \id s'~ \la \cos \theta(s) \cos \theta(s') \ra, \cr
\la y^2(t) \ra &=& v_0^2 \int_0^t \id s \int_0^t \id s'~ \la \sin \theta(s) \sin \theta(s') \ra .
\eea
To compute the position moments we first need to calculate  the auto-correlations appearing in the above equations. Let us first consider the two-time correlation of $\cos \theta;$ for $s > s'$ we have, 
\bea
C(s,s') &\equiv& \la \cos \theta(s) \cos \theta(s') \ra \cr
&=& \int \id \theta \id \theta' ~ \cos \theta \cos \theta' \cal P(\theta, s| \theta', s') \cal P(\theta, s'| 0,0 ), \n
\eea 
where the propagator $\cal P(\theta,s|\theta',s')$ satisfies the renewal equation \eqref{eq:th_renewal}. Using Eq.~\eqref{eq:th_renewal} in the above equation and performing the integrals, we get, for $s>s',$
\bea
C(s,s') &=&\frac{r^2}{(\dr+r)^2}+ \frac{2\dr}{4\dr+r} e^{-(\dr+r)s-3\dr s'} \cr
&+& \frac{r\dr}{(\dr+r)^2} \bigg(e^{-(\dr+r)s'}-e^{-(\dr+r)s}\bigg)\cr
&+& \frac{\dr^2(2\dr+5r)}{(4\dr+r)(\dr+r)^2} e^{-(\dr+r)(s-s')}.
\eea
Repeating the same exercise for $\sin \theta,$ we get,
\bea
\la \sin \theta(s) \sin \theta(s') \ra &=& \frac{2\dr}{4\dr+r} e^{-(\dr+r)(s-s')}\cr
&& \times [1- e^{-(4\dr+r)s'}].
\eea
Using the above expressions, it is straightforward to calculate the second moments. For the $x$-component we get,    
\bea
\la x^2(t) \ra &=& \frac{v_0^2}{(r+\dr)^2}\left[r^2 t^2 + \frac{2 \dr t (2\dr^2 + 9 r \dr + r^2)}{(\dr+r)(4\dr+r)} \right ] \n \\[0.25 em] 
&+& \frac{4v_0^2 e^{-(4\dr+r)t}}{3(4\dr+r)^2}  - \frac{6 v_0^2\dr^2 (2 \dr^2+16 r \dr + 5 r^2)}{(\dr+r)^4(4\dr+r)^2} \n \\[0.25 em] 
&+& \frac{2 v_0^2 e^{-(\dr+r)t}}{(\dr+r)^3} \bigg(r \dr t + \frac{4 \dr^3+33 \dr^2 r-2 r^3}{3(\dr +r)(4\dr +r)}\bigg).\n \\[0.25 em] 
&& \label{eq:x2t_th_reset}
\eea

The MSD $\sigma_{x}^2(t) = \la x^2(t) \ra - \la x(t) \ra^2 $ is then given by,

\bea 
 \sigma_{x}^2(t)  &=& \frac{2 v_0^2  \dr^2  (2\dr + 5 r) t}{(\dr+r)^3(4\dr+r)} \n \\[0.25 em]
 &+& \frac{4v_0^2 e^{-(4\dr+r)t}}{3(4\dr+r)^2} - \frac{v_0^2\dr^2 e^{-2(\dr+r)t}}{(\dr+r)^4}  \n \\[0.25 em]
 &+& \frac{4v_0^2 e^{-(\dr+r)t}}{(\dr+r)^3} \Bigg( r \dr t + \frac{8 \dr^3+ 18 \dr^2 r- r^3}{3(\dr +r)(4\dr +r)}\Bigg) \n \\[0.25 em]
&-&  \frac{v_0^2\dr^2 (28 \dr^2+104 r \dr + 31 r^2)}{(\dr+r)^4 (4\dr+r)^2}. \label{eq:varx_th_reset}
\eea


%

Similarly, we also calculate the second moment of the $y$-component,
\bea
\la y^2(t) \ra &=& \frac{4 v_0^2 \dr t}{(\dr+r)(4\dr+r)} -\frac{4v_0^2 \dr(5\dr+2r) }{(\dr+r)^2(4\dr+r)^2}\n \\[0.25 em]
&+& \frac {4 v_0^2}3 \bigg[\frac {e^{-(\dr+r)t}}{(\dr+r)^2}-\frac{e^{-(4\dr+r) t}}{(4\dr+r)^2}\bigg].\label{eq:y2t_th_reset}
\eea
The short-time and late-time behaviour of the mean-squared displacements along $x$ and $y$ are quoted in Eqs.~\eqref{eq:varxy_short_th} and \eqref{eq:Deff_th}, respectively.

\end{document}